\documentclass[12pt]{article}
\usepackage[table]{xcolor}
\usepackage[normalem]{ulem}
\usepackage{longtable}
\usepackage{footmisc}
\usepackage{orcidlink}

\usepackage{scicite}

\usepackage{aas_macros}

\usepackage{comment}
\usepackage[sectionbib]{bibunits}
\defaultbibliographystyle{sn-nature} 
\defaultbibliography{refs}

\usepackage{cancel}
\usepackage[normalem]{ulem}
\usepackage{graphicx}
\usepackage{setspace}
\usepackage{amsmath}
\usepackage{amssymb}
\usepackage{xspace}
\usepackage{multirow}
\usepackage{makecell}
\usepackage{array}
\newcolumntype{C}[1]{>{\centering\arraybackslash}p{#1}}
\usepackage{rotating}

\usepackage{siunitx}
\usepackage[flushleft]{threeparttable}

\renewcommand{\figurename}{Figure}
\renewcommand{\tablename}{Table}

\usepackage[final]{pdfpages}
\usepackage[left]{lineno}

\usepackage{times}
\usepackage[hypcap=false]{caption} 
\usepackage{subcaption, booktabs}

\usepackage{floatrow}
\DeclareFloatVCode{somespace}{\vspace{1.667\baselineskip}}
\usepackage{hyperref}

\newenvironment{sciabstract}{%
\begin{quote} \bf}
{\end{quote}}

\newcounter{lastnote}

\newcommand{\kms}{km s$^{-1}$\xspace}
\newcommand{\obj}{LEDA~145386\xspace}
\newcommand{\abw}{AT2020afhd\xspace}

\makeatletter
\renewcommand{\fnum@figure}{\textbf{Figure \thefigure}}
\renewcommand{\fnum@table}{\textbf{Table \thetable}}
\makeatother

\newcommand{\teaser}[1]{%
  \begin{center}
    \large\textit{#1}
  \end{center}
}

\newcommand{\subtitle}[1]{%
  \begin{center}
    \large\textbf{#1}
  \end{center}
}

\title{Detection of disk-jet co-precession in a tidal disruption event}

\author{
Yanan Wang$^{1\dag}$\thanks{Corresponding author. Email: wangyn@bao.ac.cn; leiwh@hust.edu.cn; huangyang@ucas.ac.cn; jfliu@nao.cas.cn. $^{\dag}$These authors contributed equally to this work.}~\orcidlink{0000-0003-3207-5237}, 
Zikun Lin$^{1,2\dag}$~\orcidlink{0000-0001-9576-1870}, 
Linhui Wu$^{3\dag}$~\orcidlink{0000-0003-3454-6522}, 
Wei-Hua Lei$^{4*}$~\orcidlink{0000-0003-3440-1526}, \\
Shuyuan Wei$^{2}$~\orcidlink{0009-0005-5669-8465}, 
Shuang-Nan Zhang$^{2,5}$~\orcidlink{0000-0001-5586-1017}, 
Long Ji$^{6}$~\orcidlink{0000-0001-9599-7285}, \\
Santiago del Palacio$^{7}$~\orcidlink{0000-0002-5761-2417}, 
Ranieri D. Baldi$^{8}$~\orcidlink{0000-0002-1824-0411}, 
Yang Huang$^{1,2*}$~\orcidlink{0000-0003-3250-2876}, \\
Ji-Feng Liu$^{1,2,9,10*}$~\orcidlink{0000-0002-2874-2706}, 
Bing Zhang$^{11,12}$~\orcidlink{0000-0002-9725-2524},
Aiyuan Yang$^{1}$~\orcidlink{0000-0003-4546-2623},  \\
Ru-Rong Chen$^{1}$, 
Yangwei Zhang$^{2}$~\orcidlink{0009-0000-7791-8192}, 
Ai-Ling Wang$^{5}$~\orcidlink{0000-0002-7351-5801}, 
Lei Yang$^{13}$~\orcidlink{0009-0002-3809-1609}, \\
Panos Charalampopoulos$^{14}$~\orcidlink{0000-0002-0326-6715},
David R. A. Williams-Baldwin$^{15}$~\orcidlink{0000-0001-7361-0246},\\ 
Zhu-Heng Yao$^{16}$~\orcidlink{0009-0000-1228-2373}, 
Fu-Guo Xie$^{3}$~\orcidlink{0000-0001-9969-2091}, 
Defu Bu$^{17}$~\orcidlink{0000-0002-0427-520X}, 
Hua Feng$^{5}$~\orcidlink{0000-0001-7584-6236},  \\
Xinwu Cao$^{18}$~\orcidlink{0000-0002-2355-3498},
Hongzhou Wu$^{4}$, 
Wenxiong Li$^{1}$~\orcidlink{0000-0002-0096-3523}, 
Erlin Qiao$^{1}$~\orcidlink{0000-0001-8319-6034},\\ 
Giorgos Leloudas$^{19}$~\orcidlink{0000-0002-8597-0756}, 
Joseph P Anderson$^{20,21}$~\orcidlink{0000-0003-0227-3451}, 
Xinwen Shu$^{13}$~\orcidlink{0000-0002-7020-4290}, \\
Dheeraj R. Pasham$^{22}$~\orcidlink{0000-0003-1386-7861}, 
Hu Zou$^{1}$~\orcidlink{0000-0002-6684-3997},
Matt Nicholl$^{23}$~\orcidlink{0000-0002-2555-3192}, 
Thomas Wevers$^{24,25}$~\orcidlink{0000-0002-4043-9400}, \\
Tom\'as E. M$\ddot{u}$ller-Bravo$^{26,27}$~\orcidlink{0000-0003-3939-7167}, 
Jing Wang$^{16,28}$, 
Jian-Yan Wei$^{16}$, \\
Yu-Lei Qiu$^{16}$, 
Wei-Jian Guo$^{1}$~\orcidlink{0000-0001-9457-0589}, 
Claudia P. Guti\'errez$^{29,30}$~\orcidlink{0000-0003-2375-2064},\\ 
Mariusz Gromadzki$^{31}$~\orcidlink{0000-0002-1650-1518},
Cosimo Inserra$^{32}$~\orcidlink{0000-0002-3968-4409},
Lydia Makrygianni$^{33}$~\orcidlink{0000-0002-7466-4868}, \\
Francesca Onori$^{34}$~\orcidlink{0000-0001-6286-1744}, 
Tanja Petrushevska$^{35}$~\orcidlink{0000-0003-4743-1679}, 
Diego Altamirano$^{36}$~\orcidlink{0000-0002-3422-0074}, \\
Llu\'is Galbany$^{29,30}$~\orcidlink{0000-0002-1296-6887}, 
Miguel Per\'ez-Torres$^{37,38}$~\orcidlink{0000-0001-5654-0266}, 
Ting-Wan Chen$^{39}$~\orcidlink{0000-0002-1066-6098}\\}
 
\begin{document}
\baselineskip24pt
\maketitle 

\subtitle{Detection of disk-jet co-precession}
\teaser{A disrupted star reveals a wobbling black hole disk and jet, seen in rhythmic X-ray and radio pulses every 19.6 days.}

\newenvironment{affiliations}{%
    \setcounter{enumi}{1}%
    \setlength{\parindent}{0in}%
    \slshape\sloppy%
    \begin{list}{\upshape$^{\arabic{enumi}}$}{%
        \usecounter{enumi}%
        \setlength{\leftmargin}{0in}%
        \setlength{\topsep}{0in}%
        \setlength{\labelsep}{0in}%
        \setlength{\labelwidth}{0in}%
        \setlength{\listparindent}{0in}%
        \setlength{\itemsep}{0ex}%
        \setlength{\parsep}{0in}%
        }
    }{\end{list}\par\vspace{12pt}}

\begin{affiliations}
\item {\small National Astronomical Observatories, Chinese Academy of Sciences, 20A Datun Road, Beijing 100101, China}
\item {\small School of Astronomy and Space Sciences, University of Chinese Academy of Sciences, Beijing 100049, China}
\item {\small Shanghai Astronomical Observatory, Chinese Academy of Sciences, 80 Nandan Road, Shanghai 200030, China}
\item {\small Department of Astronomy, School of Physics, Huazhong University of Science and Technology, Luoyu Road 1037, Wuhan 430074, China}
\item {\small Key Laboratory of Particle Astrophysics, Institute of High Energy Physics, Chinese Academy of Sciences, 19B Yuquan Road, Beijing 100049, China}
\item {\small School of Physics and Astronomy, Sun Yat-sen University, Zhuhai 519082, China}
\item {\small Department of Space, Earth and Environment, Chalmers University of Technology, SE-412 96 Gothenburg, Sweden}
\item {\small INAF - Istituto di Radioastronomia, Via P. Gobetti 101, I-40129 Bologna, Italy}
\item {\small Institute for Frontiers in Astronomy and Astrophysics, Beijing Normal University, Beijing 102206, China}
\item {\small New Cornerstone Science Laboratory, National Astronomical Observatories, Chinese Academy of Sciences, Beijing 100012, China}
\item {\small The Hong Kong Institute for Astronomy and Astrophysics, University of Hong Kong, Pokfulam Road, Hong Kong, China}
\item {\small Department of Physics, University of Hong Kong, Pokfulam Road, Hong Kong, China}
\item {\small Department of Physics, Anhui Normal University, Wuhu, Anhui 241002, China}
\item {\small Department of Physics and Astronomy, University of Turku, FI-20014 Turku, Finland}
\item {\small Jodrell Bank Centre for Astrophysics, School of Physics and Astronomy, The University of Manchester, Alan Turing Building, Oxford Road, Manchester, M13 9PL, UK}
\item {\small Key Laboratory of Space Astronomy and Technology, National Astronomical Observatories, Chinese Academy of Sciences, Beijing 100101, China}
\item {\small Shanghai Key Lab for Astrophysics, Shanghai Normal University, 100 Guilin Road, Shanghai 200234, People's Republic of China}
\item {\small Institute for Astronomy, School of Physics, Zhejiang University, 866 Yuhangtang Road, Hangzhou 310058, People’s Republic of China}
\item {\small DTU Space, National Space Institute, Technical University of Denmark, 2800 Kgs. Lyngby, Denmark}
\item {\small European Southern Observatory, Alonso de C\'ordova 3107, Casilla 19, Santiago, Chile}
\item {\small Millennium Institute of Astrophysics MAS, Nuncio Monsenor Sotero Sanz 100, Off.104, Providencia, Santiago, Chile}
\item {\small Kavli Institute for Astrophysics and Space Research, Massachusetts Institute of Technology, Cambridge, MA, USA}
\item {\small Astrophysics Research Centre, School of Mathematics and Physics, Queens University Belfast, Belfast BT7 1NN, UK}
\item {\small Space Telescope Science Institute, 3700 San Martin Drive, Baltimore, MD 21218, USA}
\item {\small Astrophysics \& Space Institute, Schmidt Sciences, New York, NY 10011, USA}
\item {\small School of Physics, Trinity College Dublin, The University of Dublin, Dublin 2, Ireland}
\item {\small Instituto de Ciencias Exactas y Naturales (ICEN), Universidad Arturo Prat, Chile}
\item {\small Guangxi Key Laboratory for Relativistic Astrophysics, School of Physical Science and Technology, Guangxi University, Nanning 530004, China}
\item {\small Institut d'Estudis Espacials de Catalunya (IEEC), Edifici RDIT, Campus UPC, 08860 Castelldefels (Barcelona), Spain}
\item {\small Institute of Space Sciences (ICE, CSIC), Campus UAB, Carrer de Can Magrans, s/n, E-08193 Barcelona, Spain}
\item {\small Astronomical Observatory, University of Warsaw, Al. Ujazdowskie 4, 00-478 Warszawa, Poland}
\item {\small Cardiff Hub for Astrophysics Research and Technology, School of Physics \& Astronomy, Cardiff University, Queens Buildings, The Parade, Cardiff, CF24 3AA, UK}
\item {\small Department of Physics, Lancaster University, Lancaster LA1 4YB, UK}
\item {\small INAF - Osservatorio Astronomico d'Abruzzo via M. Maggini snc, I-64100 Teramo, Italy}
\item {\small Center for Astrophysics and Cosmology, University of Nova Gorica, Vipavska 11c, 5270 Ajdov\v{s}\v{c}ina, Slovenia}
\item {\small School of Physics and Astronomy, University of Southampton, Southampton, Hampshire SO17 1BJ, UK}
\item {\small Instituto de Astrofísica de Andalucía (IAA-CSIC), Glorieta de la Astronom\'ia, s/n, E-18008 Granada, Spain}
\item {\small School of Sciences, European University Cyprus, Diogenes Street, Engomi, 1516 Nicosia, Cyprus}
\item {\small Institute of Astronomy, National Central University, 300 Jhongda Road, 32001 Jhongli, Taiwan}
\end{affiliations}

\begin{bibunit}
\begin{sciabstract}
Theories and simulations predict that intense spacetime curvature near black holes bends the trajectories of light and matter, driving disk and jet precession under relativistic torques.
However, direct observational evidence of disk-jet co-precession remains elusive. Here, we report the most compelling case to date: a tidal disruption event (TDE) exhibiting unprecedented 19.6-day quasi-periodic variations in both X-rays and radio, with X-ray amplitudes exceeding an order of magnitude. The nearly synchronized X-ray and radio variations suggest a shared mechanism regulating the emission regions. We demonstrate that a disk-jet Lense-Thirring precession model successfully reproduces these variations while requiring a low-spin black hole.
This study uncovers previously uncharted short-term radio variability in TDEs, highlights the transformative potential of high-cadence radio monitoring, and offers profound insights into disk-jet physics.
\end{sciabstract}

\noindent \textbf{\large Introduction}\\
Tidal disruption events (TDEs) are transient phenomena that occur when a star ventures too close to a supermassive black hole (SMBH) and is torn apart by its tidal forces (e.g., \cite{Rees1988,Evans1989}). The resulting stellar debris falls back toward the SMBH, forming a nascent accretion disk and, in some cases, launching (mildly) relativistic jets (e.g., \cite{Levan2011,Bloom2011,Cenko2012,Brown2015,Andreoni2022}). This process unfolds over timescales of months to years, offering a unique opportunity to study accretion and jet-launching physics around SMBHs in real time.


In TDEs, the angular momentum of the accretion disk, imparted by the disrupted star, is often misaligned with the spin axis of the central Kerr black hole (BH).
This misalignment is expected to induce Lense-Thirring (LT, \cite{LenseThirring1918}) precession of the disk and associated jet, driven by frame-dragging effects in the strong-field regime of general relativity (e.g., \cite{Stone2012,Franchini2016}).
General relativistic magnetohydrodynamic (GRMHD) simulations support this picture, predicting coupled disk-jet precession (e.g., \cite{Liska2018, Chatterjee2020}); however, direct observational confirmation remains elusive. Thus far, disk or jet precession has only been observed separately in various accreting systems, as evidenced by either X-ray (e.g., \cite{Saxton2012, Ma2021, Pasham2024}) or radio observations (e.g., \cite{Cui2023, Tian2023}).

The detection of disk-jet co-precession has been impeded by several observational challenges, including the transient nature of the disk and jet, limited cadence in radio monitoring, viewing angle effects, and contamination from other sources of variability. Here, we present the most compelling evidence to date for disk-jet co-precession in a recent TDE, enabled by exceptionally dense temporal coverage in both X-ray and radio observations.

\begin{figure}[htp!]
\centering
\includegraphics[width=1\textwidth]{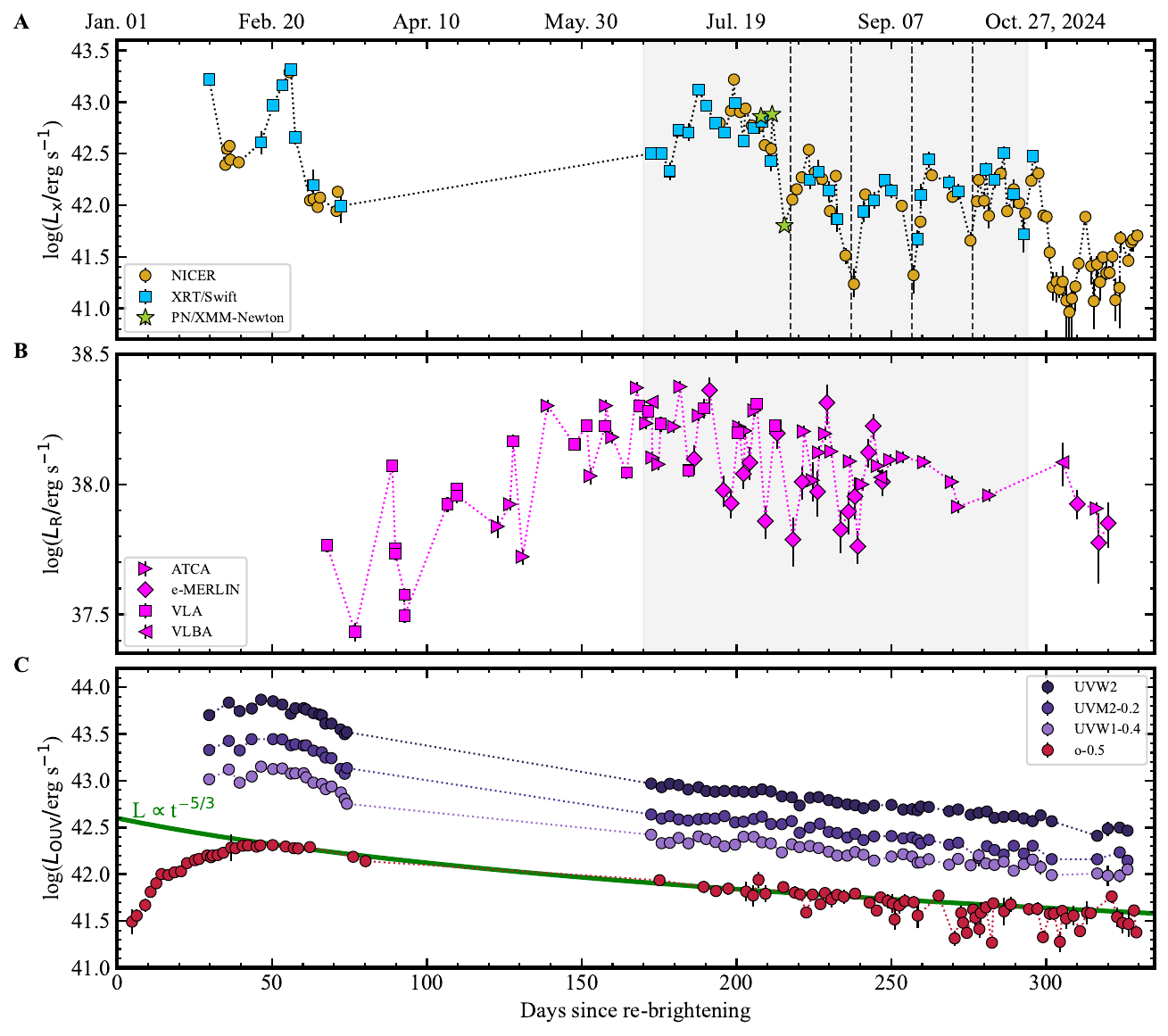}
\caption{\small \textbf{Temporal evolution of the multiwavelength luminosity of \abw since its optical re-brightening in 2024 (MJD 60310).}
(\textbf{A}) The unabsorbed X-ray (0.3--2 keV) luminosity. 
(\textbf{B}) The radio (5--6~GHz) luminosity.
The gray-shaded region represents the period used for calculating the cross-correlation function between X-ray and radio data. 
(\textbf{C}) The UV and optical luminosities.
The UVOT data were corrected for Galactic extinction and had the host contribution subtracted, while the ATLAS data were corrected for extinction. The lightcurves are offset as indicated in the legend for clarity. The green line indicates a powerlaw of $t^{-5/3}$. Uncertainties are quoted
at the 1$\sigma$ confidence level.}\label{fig:lc}
\end{figure}

\vspace{1cm}
\noindent \textbf{\large Results}\\
\vspace{0.3cm}
\noindent \textbf{Optical counterpart}\\
AT2020afhd (a.k.a. ZTF20abwtifz) is an optical transient situated at the nucleus of the galaxy \obj (fig.~\ref{fig:image_multi}), at a redshift of 0.027 \cite{Hammerstein2024,Arcavi2024 }. The transient was discovered by the Zwicky Transient Facility (ZTF) in 2020, with initial detections at a $g$-band magnitude of $\sim20$. On January 4, 2024, ZTF detected a significant re-brightening \cite{Fremling2024}, with a peak magnitude of $g_{\rm ABmag} \sim 16.6$. Follow-up optical spectra reveal the presence of a blue continuum (consistent with the optical colors of $g-r\sim -0.02$), an He {\sc ii} emission line and broad Balmer emission, leading to the classification of the re-brightening as a TDE \cite{Hammerstein2024}. 
Moreover, the optical decline rate follows approximately a $t^{-5/3}$ power-law (see Fig.~\ref{fig:lc}), consistent with the debris fallback rate predicted by TDE theories \cite{Rees1988,Phinney1989,Evans1989}.
Additionally, we determined the central BH mass using two independent methods (Supplementary~\ref{sec:bhmass}), both of which yielded consistent results. For the subsequent analysis in this work, we adopt a BH mass of $\log(M_{\rm BH}/M_{\odot})=6.7\pm0.5$ and take MJD~60310 as the initial date of the re-brightening.

\vspace{0.3cm}
\noindent \textbf{X-ray counterpart}\\
In X-rays, several Neil Gehrels Swift Observatory (Swift) monitoring programs commenced since January 26, 2024 (25 days after the re-brightening). These observations revealed significant variations on timescales of approximately 25--40 days, during which the peak luminosities exceeded the dips by more than one order of magnitude (see Fig.~\ref{fig:lc}). Such variations are also evident in our Neutron star Interior Composition ExploreR (NICER) campaign, triggered 10 days after the initial Swift program. The peak X-ray luminosity is approximately two orders of magnitude higher than in 2020, suggesting the variations are associated with the newly occurred TDE. Moreover, during the first 300 days, the X-ray spectrum remains ultrasoft, dominated by a multi-color disk blackbody component with $kT_{\rm in}\sim10^{5.7-6.4}~$K.
The derived blackbody temperature closely follows the evolution of the X-ray luminosity, increasing with rising luminosity and decreasing as it declines (see Fig.~\ref{fig:spec}A).

\begin{figure}[htbp!]
\centering
\includegraphics[width=1\textwidth]{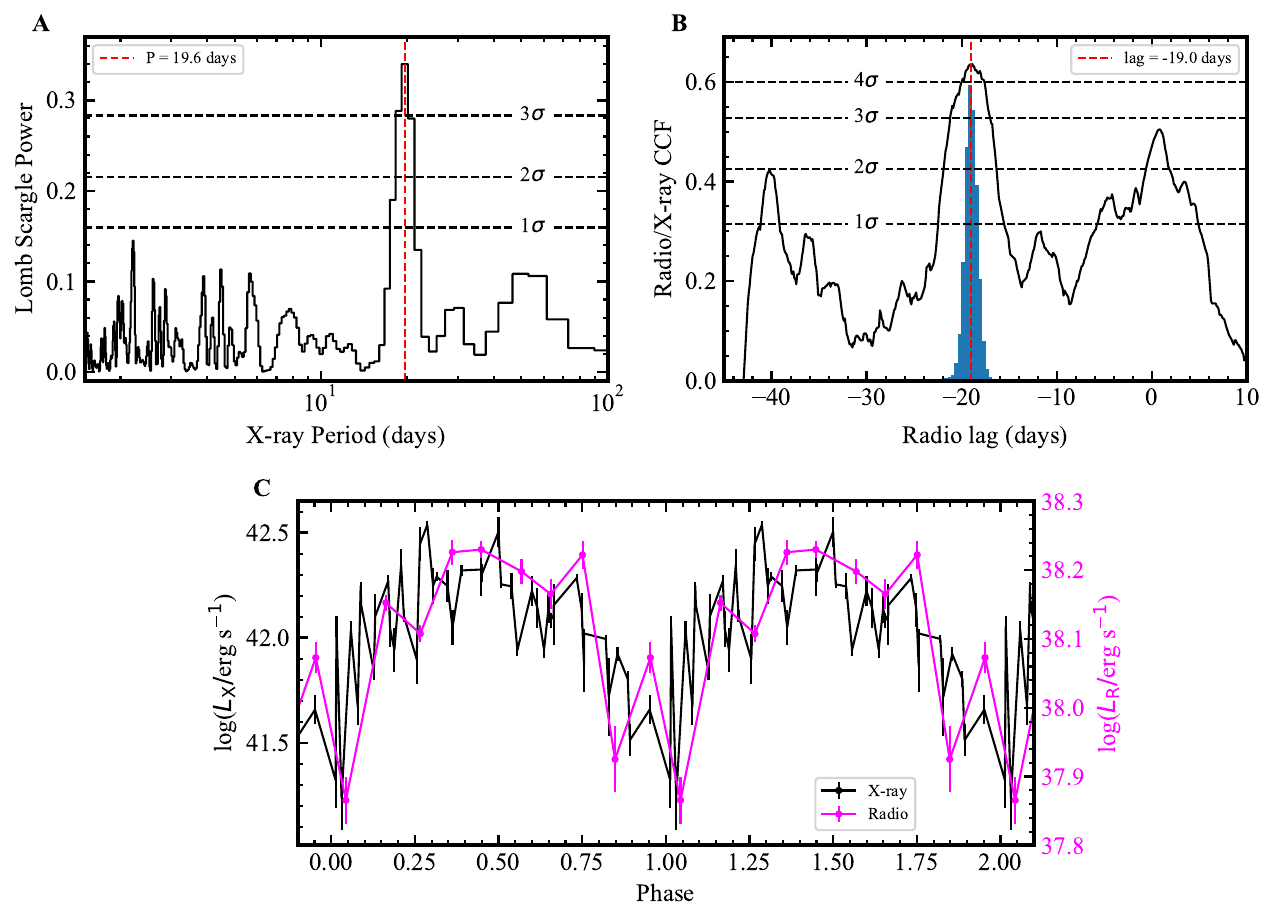} 
\caption{\small \textbf{Timing analysis of the X-ray and radio data.}
(\textbf{A}) Lomb-Scargle periodogram of the X-ray lightcurve. The calculation includes data collected between August 3 and October 21, 2025, during which the X-ray quasi-periodic variations were clearly observed. (\textbf{B}) Cross-correlation function between the X-ray and radio data. The data used in this calculation are indicated by the gray-shaded region in Figs.~\ref{fig:lc}A and B. The histogram represents the distribution obtained from bootstrap simulations, with the red dashed line marking the median of the distribution at $-19.0$ days. 
(\textbf{C}) Folded X-ray and radio lightcurves with a period of 19.6 days. The radio data were rebinned into 0.1-phase intervals using a weighted mean for clarity. The data used in this calculation are shown in Fig.~\ref{fig:schematics_model}B.}\label{fig:timing}
\end{figure}

By 215 days after the re-brightening, the 0.3--2 keV X-ray luminosity had dropped by over an order of magnitude and began exhibiting clear periodic variations, visible to the naked eye. The variation amplitude remained above an order of magnitude.  
A Lomb-Scargle periodogram (LSP) of background-subtracted X-ray data from 215--294 days (August 3 to October 21) identifies a period of $19.6 \pm 1.5$ days with a statistical significance exceeding $3.79\sigma$ (Fig.~\ref{fig:timing}A, see Materials and Methods for details). Notably, the short-term modulation is absent in the high-cadence optical and UV photometry.
The consistent periodic behavior observed in both the luminosity and blackbody temperature is reminiscent of the TDE AT2020ocn \cite{Pasham2024}. In a similar case, AT2020ocn exhibited X-ray modulations with a 15-day period lasting approximately 130 days, which were attributed to LT precession, although the possibility of radiation-pressure instabilities could not be entirely excluded.

\vspace{0.3cm}
\noindent \textbf{Radio counterpart}\\
In radio bands, {\abw} was detected as a point source with the Karl G. Jansky Very Large Array (VLA) three days after the first X-ray detection, with a flux density of $253\pm14~\mu$Jy at 15.1 GHz \cite{Christy2024}. 
The host galaxy of {\abw} did not show any past radio activity prior to the re-brightening. Hence, the nascent radio counterpart is also very likely associated with the TDE. 
Approximately 67 days later, we initiated high-cadence radio monitoring of {\abw} at C band using the VLA, the Australia Telescope Compact Array (ATCA), the Enhanced Multi-Element Remotely Linked Interferometer Network (e-MERLIN), and the Very Long Baseline Array (VLBA). Remarkably, this comprehensive campaign uncovers radio variations with a period of approximately 20$-$40 days and a peak-to-dip ratio of exceeding four (see Fig.~\ref{fig:lc}). 
Besides that, we calculated the in-band photon index, $F \propto \nu^{\alpha}$, using VLA and ATCA data and determined the peak frequency, $\nu_{\rm p}$, and peak flux density, $F_{\rm p}$, of the radio broadband spectral energy distribution (SED) using VLA data. Both the long-term increase in the 5-GHz luminosity and the transition of $\alpha$ from positive to negative values suggest that the spectrum evolved from optically thick to optically thin, consistent with the evolution of the SED (see Figs.~\ref{fig:spec}B and C, and Supplementary~\ref{sec:radio_sed}).

\begin{figure}[htbp!]
\centering
\includegraphics[width=1\textwidth]{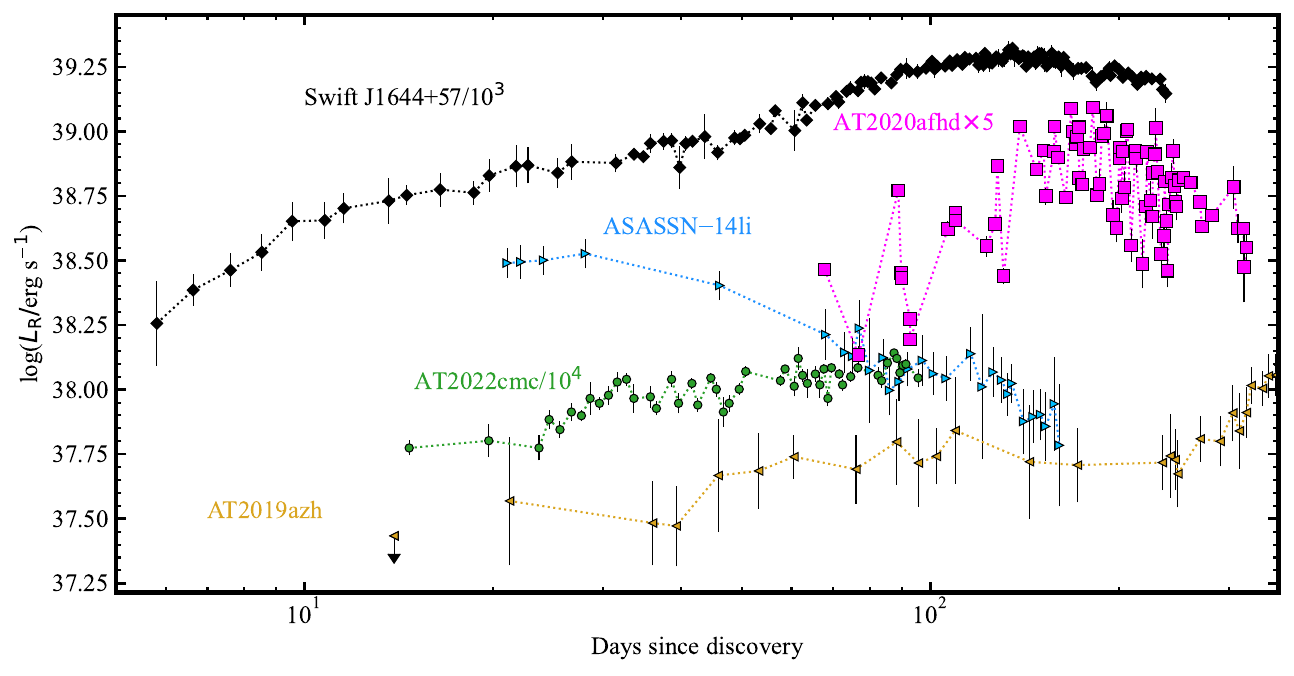}
\caption{\small \textbf{Comparison of TDEs with early intense radio detection.} The luminosities of the two on-axis jetted TDEs, Swift~J1644+57 and AT2022cmc, as well as \abw, have been rescaled for clarity.
The AMI-LA data at 15.5~GHz for Swift~J1644+57, ASSASN--14li, AT2019azh and AT2022cmc were adapted from \cite{Berger2012}, \cite{Bright2018}, \cite{Sfaradi2022}, and \cite{Rhodes2023}, respectively.}\label{fig:radio_TDEs}
\end{figure}

Compared to other TDEs with early radio detections (within a hundred days of discovery), \abw exhibited unprecedented high-amplitude variations in radio bands on short-term timescales (as shown in Fig.~\ref{fig:radio_TDEs}), pointing to the emergence of a new class of radio TDEs. 
Our VLBA U- and C-band observations across different epochs confirmed that the emitting region remained point-like (fig.~\ref{fig:vlba}). The consistent flux densities observed with VLBA, VLA, and e-MERLIN at similar times indicate that the radio emission, at least at 5--6 GHz, originated from a compact, unresolved source, effectively ruling out angular resolution effects. 
Flux calibration offsets among the VLA, e-MERLIN and ATCA observations, expected to be up to 22\% (Supplementary~\ref{sec:radioflux_offset}), cannot account for the observed variations. The effect of interstellar scintillation \cite{Walker1998MNRAS}, which would produce hour-scale variations, is excluded based on the consistent flux densities detected in multiple VLA observations taken within one hour (Fig.~\ref{fig:lc}B).
Emission resulting from winds or jet-wind interactions with the circumnuclear medium is expected to be isotropic and should not exhibit significant modulation on timescales of tens of days. Therefore, the previously uncharted short-term radio variations observed in \abw are most likely driven by the dynamic evolution of jets.

\vspace{0.3cm}
\noindent \textbf{Cross-correlation between X-ray and radio emission}\\
Unfortunately, the sparse sampling of radio observations, combined with contamination from non-periodic variations on both long timescales (hundreds of days) and short timescales (days), makes it challenging to directly extract periodicity from the radio data and reliably assess its significance through comprehensive analysis.  
Instead, we evaluated the connection between X-ray and radio variations using the discrete cross-correlation function (\cite{Sun2018}, see Materials and Methods). 
This analysis revealed a significant $4.26\sigma$ correlation between the X-ray and radio emissions, with a primary peak at a time lag of $-19.0^{+0.7}_{-0.6}$ days (Figs.~\ref{fig:timing}B and C). 
In addition, two secondary peaks are found at lags of 0 and $-40$ days, corresponding to integer multiples of the X-ray variability period. 
These findings suggest that the primary lag of $-19$ days may result from the limited temporal coverage of the data and noise across different timescales, rather than indicating a definitive physical lead of the radio emission. 
The nearly synchronized radio and X-ray variations suggest a common mechanism tightly regulating both emissions. 
The tens-of-day period aligns with disk precession in TDEs \cite{Stone2012,Franchini2016}, making synchronized disk-jet precession the most natural explanation.

\vspace{0.3cm}
\noindent \textbf{Disk-jet co-precession model}\\
To test the disk-jet precession scenario, we first introduced a rigid-body LT precession model under conditions where
the accretion rate remained above the Eddington limit during the first year of a TDE involving a SMBH with a mass of $10^{6.7\pm0.5}~M_{\odot}$ disrupted a solar-like star. During this period, the disk remains geometrically thick, characterized by a disk angular semi-thickness $H/R>\alpha$, where $\alpha$ is the disk viscosity parameter \cite{Shakura1973}, $H$ is the disk thickness, and $R$ is the disk radius. The disk is assumed to extend from the ISCO to the debris circularization radius \cite{Stone2012,Franchini2016}, and the amplitude of the X-ray variations is attributed to changes in the disk's projected area and periodic obscuration by the outer disk (see Fig.~\ref{fig:schematics_model}A and Supplementary~\ref{sec:disk_precession}).
The BH spin parameter can then be estimated by assuming that the observed period $T_{\rm prec}\sim19.6$~days corresponds to the LT precession period.
Adopting several power-law indices for the surface density profile \cite{Franchini2016}, and a BH mass range of $M_\bullet=10^{6.7\pm0.5}~M_\odot$, we found that the spin parameter $a_\bullet$ fell within the range of $-(0.46-0.14)$ or $0.11-0.35$ (Fig.~\ref{fig:MCMC_precession}A). Since a negative spin parameter would result in a larger ISCO and, consequently, a larger inner disk radius, it would be challenging to explain the observed hot disk. Therefore, a positive spin parameter was favored. 
By modeling the X-ray lightcurve, we constrained the observer's viewing angle relative to the BH spin to \(\theta_{\rm obs} \sim 38.4 ^{+0.5\circ}_{-0.6}\), and the disk precession angle to \(\theta_{\rm i} \sim 14.5^\circ \pm0.5^\circ\) (Fig.~\ref{fig:MCMC_precession}B).

We then examined the jet precession scenario, in which the radio luminosity varies as the jet cone shifts toward and away from the line of sight. The peak-to-dip ratio is determined by the Doppler factor ratio, which depends on the jet Lorentz factor (\(\Gamma\)) and the angle between the observer and the jet axis (Supplementary~\ref{sec:jet_precession}). 
The luminosity peaks when the difference between $\theta_{\rm obs}$ and $\theta_{\rm i}$ is minimized and dips when it is maximized. Adopting $\theta_{\rm obs} \sim 38.4^\circ$ and $\theta_{\rm i} \sim 14.5^\circ$, we derived $1.2 \leq \Gamma \leq 1.6$. 
Jets can be powered by two primary mechanisms: the Blandford-Znajek (BZ; \cite{Blandford1977}) mechanism, driven by BH spin, and the Blandford-Payne (BP; \cite{Blandford1982}) mechanism, powered by disk rotation. 
Both mechanisms can account for our radio observations, with the BZ mechanism requiring a stronger magnetic field than the BP mechanism, i.e., $B_{\bullet} \sim 2.8 \times 10^{3}\rm~G$ for the former and $B_{\bullet} \sim 1.5 \times 10^{3}\rm~G$ for the latter, assuming a Lorentz factor of $ \Gamma = 1.6$. In Fig.~\ref{fig:schematics_model}B, we present a comparison between our disk-jet precession model and the observations.

Around 250 days after the re-brightening, the timescale of radio variations appeared to lengthen, while the X-ray variation profile remained unchanged. After 300 days, the X-ray emission dropped rapidly, and the 19.6-day quasi-periodic variations disappeared. Meanwhile, the radio emission weakened and became anti-correlated with the X-ray emission. Soon after, the UV-optical emission exhibited a second re-brightening, which we exclude from this study to maintain focus.
The radio sampling after 250 days is insufficient for detailed tracking of its co-variances with X-rays, but the disk-jet connection clearly broke around 300 days. Current GRMHD simulations explore disk-jet co-precession only over limited timescales, predicting a gradual slowdown of both precession and alignment \cite{Liska2018,Chatterjee2020}. However, when and how this connection breaks, along with the subsequent independent evolution of the components, remain unexplored, warranting future theoretical investigations.

\vspace{1cm}
\noindent \textbf{\large Discussion}\\
\noindent \textbf{Other potential mechanisms driving X-ray and radio co-variances}\\
Disk radiation pressure instabilities may also produce comparable periodicity in X-ray and radio emissions. For example, the microquasar GRS~1915+105 exhibited simultaneous periodic modulations in X-ray and radio bands on timescales of 20--50 minutes \cite{Pooley1997, Mirabel1998, Vincentelli2023}. This periodic flaring activity has been attributed to radiation pressure instabilities, where material in the inner region of an optically thick accretion disk is rapidly depleted and replenished \cite{Belloni1997a, Belloni1997b}. During this process, part of the inner disk is ejected to form a jet, as indicated by a rise in the radio flare coinciding with a dip in the X-ray flare \cite{Pooley1997}. However, the X-ray and radio variations in GRS~1915+105 are considerably more complex, resulting in diverse correlations between the two-band lightcurves \cite{Klein2002}. Overall, the lack of simulations and theoretical models on the short-term evolution of jets during radiation-pressure instabilities makes it challenging to further test this scenario.

Although the correlation between the X-ray and radio variations of \abw suggests that the latter is unlikely to have an external origin, we investigated the potential effect of interstellar scintillation (ISS, \cite{Walker1998MNRAS}) on our observations. Using the NE2001 model \cite{Cordes2002}, we calculated the transition frequency between the strong and weak scintillation regimes to be $\nu_0 = 7.81$ GHz, indicating that our C-band observations ($\sim6\rm~GHz$) fall within the strong, refractive scintillation regime.
According to \cite{Walker1998MNRAS}, the modulation index and timescale are given by $m_{\rm p} = (\nu/\nu_0)^{17/30}= 0.86$ and $t_{\rm r} = 2(\nu_0/\nu)^{11/5}= 3.6$ hrs, respectively. 
In our VLA dataset, several pairs of observations taken within an hour showed consistent flux densities, indicating that ISS does not significantly contribute to the observed radio variations.

\noindent \textbf{Nature of the 2024 re-brightening}\label{sec:2024re-brightening}\\
\noindent \abw was initially discovered as an optical transient by the ZTF on October 20, 2020 \cite{Bellm2019}. This event could be associated with nuclear activity in the host galaxy (2MASX~J0313357--020907), which transitioned from Seyfert II \cite{Chen2022} to Seyfert I. 
Supporting evidence includes the evolution of the H$\alpha$ emission line, which appeared narrow and weak in the 6dF spectrum from 2005, but became broad and prominent in the DESI spectrum from 2022, with a measured width of $\sigma = 3298 \pm 81~\mathrm{km\,s^{-1}}$.
Additionally, weak X-ray activity was observed, with a luminosity of $10^{41.7 \pm 0.8} \rm erg\,s^{-1}$ detected by eROSITA-SRG on February 7, 2020 (fig.~\ref{fig:xray_spectra}D).
In the radio band, AT2020afhd was not detected in the FIRST 1.4 GHz catalog (with an upper limit of $<$0.54 mJy beam$^{-1}$) or in the VLASS survey at 3 GHz during three epochs (November 2017, September 2020, and March 2023). The rms noise levels for the individual VLASS observations were approximately 0.15 mJy beam$^{-1}$, and the mosaicked image had a sensitivity limit of $<$0.08 mJy beam$^{-1}$.

\begin{figure}[htbp!]
\centering
\includegraphics[width=1\textwidth]{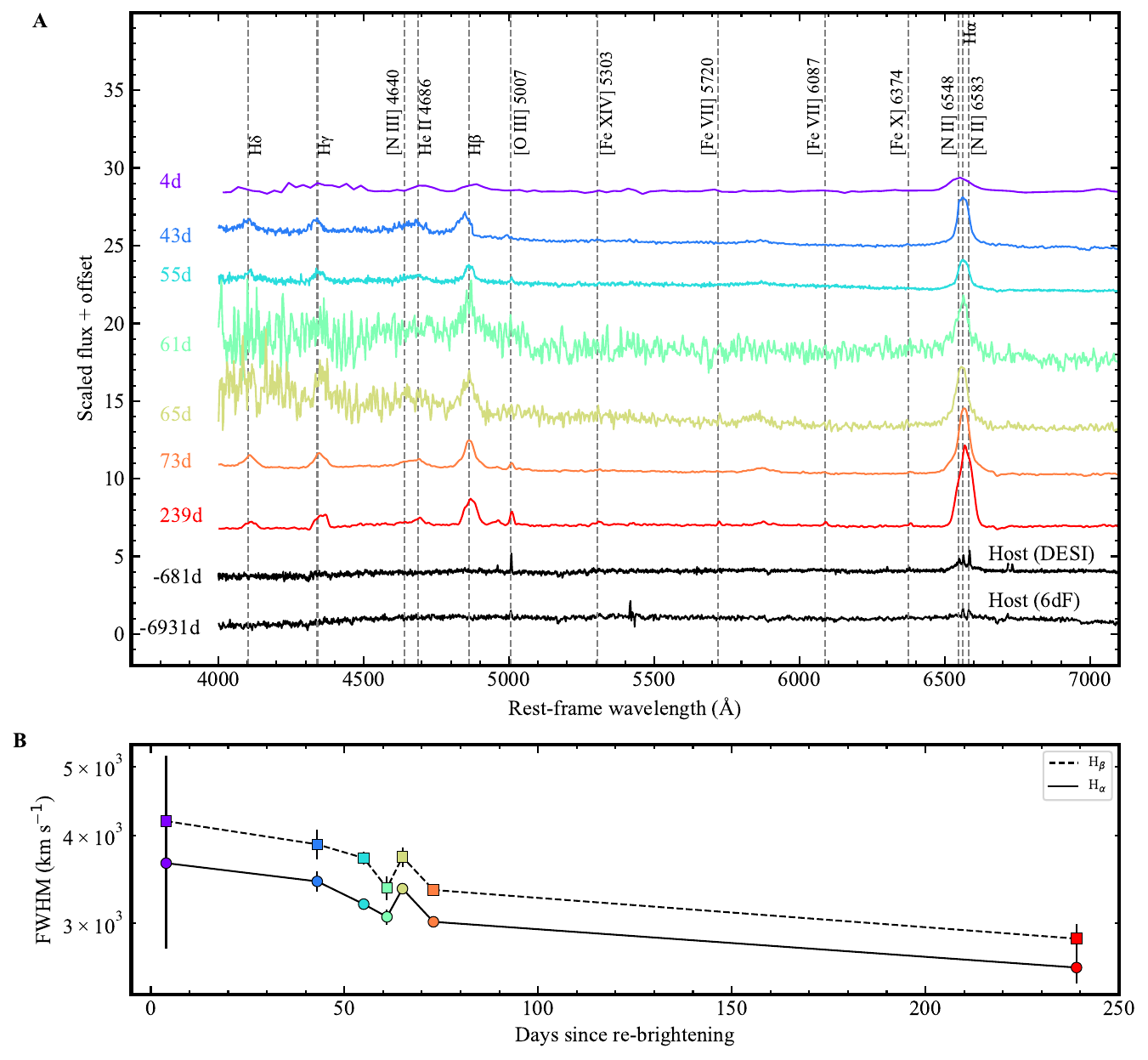}
\caption{\small \textbf{Optical spectra of AT2020afhd.} (\textbf{A}) The spectroscopic evolution of AT2020afhd. The spectra have been re-scaled and offset for clarity, with rest-frame phases indicated. (\textbf{B}) Evolution of the FWHM of the H$\alpha$ and H$\beta$ emission lines.}\label{fig:opt_spectra}
\end{figure}

After three and a half years (beginning in 2024), the source showed signs of re-brightening, starting at $g_{\rm AB}\sim19.5$ on January 5, 2024, peaking at $g_{\rm AB}\sim16.8$ on February 10, and then fading to $g_{\rm AB}\sim18.4$ on December 22 (see fig.~\ref{fig:host_info}A). It was initially classified as a TDE by \cite{Hammerstein2024} based on its persistent blue optical colors, strong UV flux, broad Balmer emission, and broad He {\sc ii} emission. Later, \abw was reclassified as a Bowen fluorescence flare (BFF, \cite{Arcavi2024}), primarily based on the widths of its He {\sc ii} and Balmer emission lines. 
However, unlike the relatively stable long-term optical, UV and X-ray emissions observed in BFFs \cite{Trakhtenbrot2019}, \abw showed a significant decline, nearly 2 magnitudes in UV photometry and over two orders of magnitude in X-rays, over the course of about a year. Combined with its thermal X-ray spectrum, this behavior aligns more closely with that typically seen in TDEs. Additionally, it remains unclear whether BFFs are triggered by TDEs.
We summarized the observation log for the optical spectra with an SNR greater than 10 in Table~\ref{tab:opt_sum} and showed the spectra in Fig.~\ref{fig:opt_spectra}A. 
Fig.~\ref{fig:opt_spectra}B illustrates the evolution of the full width at half maximum (FWHM) of H$\alpha$ and H$\beta$, which decreases as the luminosities decrease.

Similar to AT2020ocn \cite{Pasham2024}, the optical-UV lightcurves of \abw were dominated by a long-term declining trend with no significant short-term variations (see Fig.~\ref{fig:lc} and fig.~\ref{fig:host_info}A), whereas the X-ray and radio photons exhibited high variability on timescales of tens of days.
These distinct optical-UV behaviors, compared to radio and X-rays, could be explained by X-ray reprocessing \cite{Guillochon2013, Roth2016} or stream-stream collisions \cite{Bonnerot2020}. In either scenario, the optical-UV emission was produced at a significant distance from the central engine.

In the latest NTT spectrum (2024-08-27, at +239 days), we detected high-ionization coronal lines (CLs). These include Fe {\sc xiv} $\lambda5303$, Fe {\sc vii} $\lambda\lambda 5720,6087$, and Fe {\sc x} $\lambda6374$. There are indications of CLs in the first NTT spectrum taken before the seasonal gap, although they appear substantially weaker.
CLs are high-ionization lines (e.g., of Iron, Neon, Argon, Sulfur) that originate from the photoionization or collisional ionization of a clumpy interstellar medium or other pre-existing material. Although typically associated with AGN, strong CLs have been detected in the optical spectra of some galaxies that show little to no evidence of AGN activity \cite{Komossa2008,Wang2012}. Many of these CLs have ionization potentials of $\geq 100$ eV, requiring a strong extreme UV (EUV) and/or soft X-ray ionizing continuum. It has long been thought that such extreme coronal line emitters (ECLEs) are powered by TDEs, and recent studies support this scenario \cite{Hinkle2024,Clark2024}. Interestingly, a few optical-UV TDE candidates \cite{Onori2022,Short2023} out of the $\sim50$ strong candidates discovered to date have exhibited late ($\geq 200$ days post-peak) X-ray brightening accompanied by the emergence of iron coronal lines. 
In contrast, in the case of \abw, the CLs appear to emerge significantly later than the onset of X-ray emission. Further investigation, including detailed modeling, is required to understand the origin of the CLs, which will be explored in a forthcoming study.

In summary, \abw exhibits unprecedented high-amplitude, synchronized quasi-periodic variations in X-ray and radio bands, providing the first known evidence that the disk and jet can co-precess on comparable timescales.
TDEs, with evolution timescales spanning hundreds of days, serve as unique laboratories for studying the dynamics of nascent accretion disks and jets. Building on the case of AT2020afhd, we propose using modulated X-ray variations as triggers for high-cadence radio follow-ups. This approach aims to efficiently expand the sample of such TDEs and eventually deepen our understanding of disk-jet physics.

\vspace{1cm}
\noindent \textbf{\large Materials and Methods}\\
\noindent \textbf{Data summary}\\
\noindent \abw was well detected by both ground- and space-based facilities, with observations spanning from radio to soft X-rays (see the target's localization in fig.~\ref{fig:image_multi}). Around late November 2024, \abw exhibited another re-brightening across multiwavelengths at the time of writing. In this study, we focused exclusively on the period between January 1 and November 26, 2024, prior to the onset of the second re-brightening. Details of the observations, data reduction procedures, and X-ray spectral analysis are provided in Supplementary~\ref{sec:data} and \ref{sec:xspec}.
We adopted a flat $\Lambda$CDM cosmology with $H_{0}=67.4 \rm ~km~s^{-1}Mpc^{-1}$ and $\Omega_{\rm m}=0.315$ from \cite{Planck2020}, where a redshift of 0.027 corresponds to a luminosity distance of approximately 123~Mpc.

\vspace{0.3cm}
\noindent \textbf{Lomb-Scargle periodogram and cross-correlation function}\\
\noindent The X-ray lightcurve was derived from the unabsorbed flux of the \texttt{diskbb} component in the 0.3--2~keV band, revealing clear quasi-periodic variations between August 3 and October 21, 2024, spanning a total of 79 days. To analyze these variations, we computed the LSP \cite{Scargle1982} using the {\sc astropy} library \cite{Astropy2013} with data from this 79-day period.
To more accurately determine the period of the variations, we divided the NICER observations into individual GTIs with durations exceeding 300 seconds. The flux for each NICER GTI was measured following the method described in Supplementary~\ref{sec:data}. Our analysis was performed on a GTI basis for the NICER data and on an observation basis for the XRT and PN data.  
Given the average sampling interval of 1 day and the total duration of 79 days, 
we conducted the LSP analysis over a period search range of 1 to 100 days, similar to the range used in the study of AT2020ocn, which had a period of 15 days \cite{Pasham2024}.  
This analysis identified a period of 19.6 days with a FWHM of 3.0 days, as illustrated in the Fig.~\ref{fig:timing}A.

To assess the statistical significance of the detected period, we first quantified the contribution of the LSP continuum by modeling it as a power-law, $P(f) \propto f^{\alpha}$. The fit was applied to the observed LSP, excluding the period range corresponding to the detected signal, as defined by its FWHM. The best-fitting power-law index was determined to be $\alpha = -0.02 \pm 0.05$. This value is consistent with $0$ within the $1\sigma$ uncertainty range, indicating that the continuum was not significantly affected by red noise but was instead dominated by white noise.

\begin{figure}[htbp!]
\centering
\includegraphics[width=1\textwidth]{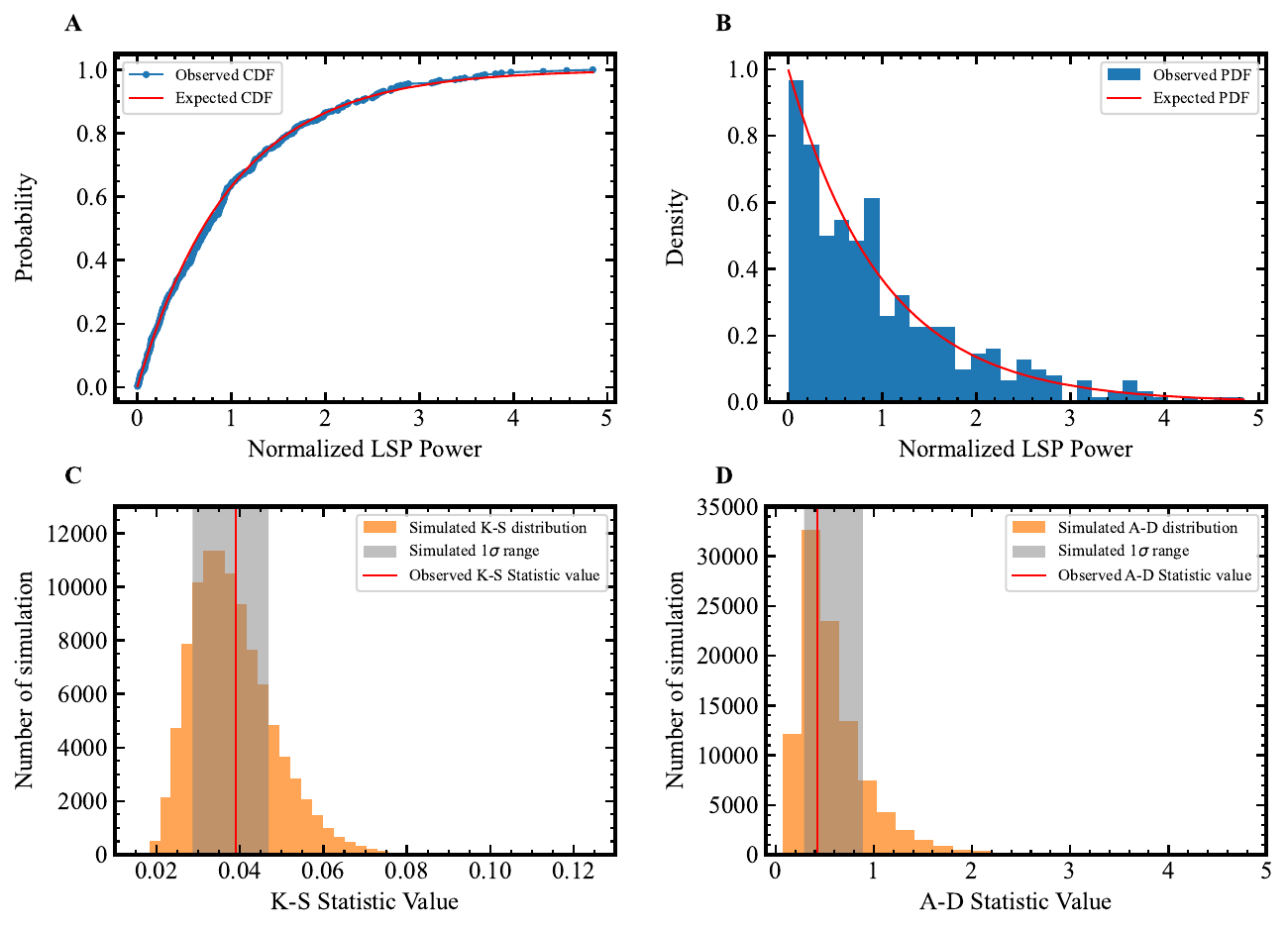}
\caption{\small \textbf{White noise tests for LSP.}
(\textbf{A}) The CDF of the normalized LSP noise power from the observed data (blue points) compared to the expected CDF of an exponential distribution (red curve).
(\textbf{B}) The PDF of the normalized LSP noise power from the observed data (blue histogram) compared to the expected PDF of an exponential distribution (red curve).
(\textbf{C}) K-S test results from simulations. The orange histogram shows the distribution of K-S statistic values for simulated white noise, with the $1\sigma$ range shaded in grey. The red line represents the K-S statistic value for the observed LSP noise power. 
(\textbf{D}) A-D goodness-of-fit test results from simulations. The orange histogram shows the distribution of A-D statistic values for simulated white noise, with the $1\sigma$ range shaded in grey. The red line represents the A-D statistic value for the observed LSP noise power.}\label{fig:KSAD_test}
\end{figure}

To further test whether the continuum was consistent with white noise, we applied the algorithms described in \cite{Pasham2024SciA}. We compared the cumulative distribution function (CDF) and probability density function (PDF) of the LSP power values to those expected for white noise. The expected CDF followed $1 - \exp(-z)$ \cite{Scargle1982}, where $z$ represented LSP powers. The comparisons of the observed and expected CDF and PDF are shown in Figs.~\ref{fig:KSAD_test}A and B.
To quantify these comparisons, we conducted Kolmogorov-Smirnov (K-S) and Anderson-Darling (A-D) goodness-of-fit tests on the observed and expected CDF using the {\sc scipy} library \cite{Virtanen2020}. 
The K-S test yielded a statistic of 0.0391 (p-value = 0.93), and the A-D test returned a statistic of 0.4278, which is well below the 10\% critical value of 1.07. These results indicate that, at a 90\% confidence level, the null hypothesis that the LSP power followed the expected white noise distribution could not be rejected.
Additionally, we performed 100,000 Monte Carlo simulations of the LSP continuum under the assumption of white noise and computed their corresponding K-S and A-D statistics. The distributions of these statistics are presented in Figs.~\ref{fig:KSAD_test}C and D. The K-S and A-D statistic values of the observed CDF fell within the $1\sigma$ range of the distribution of the simulated white noise statistic values. These findings robustly indicated that the LSP continuum was statistically consistent with white noise.

To determine the global statistical significance of the period, we conducted a false alarm probability (FAP) analysis. We generated 100,000 simulated lightcurves with the same temporal sampling as the observed data, allowing the flux values to vary randomly within the observed range, bounded by the minimum and maximum flux. From these simulations, we identified 15 occurrences of spurious periodic signals, corresponding to a statistical significance of approximately $3.79\sigma$.

To investigate the correlation between radio and X-ray variations of \abw, we used the Python-based discrete cross-correlation function (PyCCF, \cite{Sun2018}), focusing on data collected between June 19 and October 21, 2024. The analysis revealed that the cross-correlation function (CCF) peaked at a time lag of $\sim-19.0$ days (see Fig.~\ref{fig:timing}B). 
To assess the global significance of this correlation, we generated 100,000 random radio lightcurves, preserving the same observation sampling while randomizing the luminosities within the observed minimum and maximum values. We tested the CCF across time lags spanning $-45$ to $10$ days, a range covering $\pm 1$ cycle of the periodic signal. Among these simulations, only 2 spurious CCF signals were detected, corresponding to a statistical significance of approximately $4.26\sigma$. 
Furthermore, we estimated the uncertainties in the time lag by focusing on the range of $-30$ to $-10$ days, corresponding to a single period cycle around the CCF peak. Using 100,000 Monte Carlo simulations with a bin size of 0.1 days and bootstrapping methods to estimate uncertainties \cite{Peterson1998,Sun2018}, we determined a time lag of $-19.0^{+0.7}_{-0.6}$ days for the radio relative to the X-ray. This result is consistent with the observed period, indicating that the X-ray and radio emissions were synchronized. 
As described in Section 1.3.5, we applied a 22\% offset to the observed ATCA flux density. Using the original ATCA flux density and following the same methodology outlined above, we obtained a statistical significance of $3.85\sigma$ and a time lag of $-19.1^{+0.6}_{-0.7}$ days. These results indicate that the calibration offset had no substantial impact on the derived time lag or the statistical significance of the correlation.

Additionally, we folded the X-ray and radio lightcurves with a 19.6-day period. For clarity, the folded radio lightcurve was rebinned into 0.1-phase intervals using a weighted mean, with error estimation described in \cite{Standing2023NatAs}. The folded X-ray and rebinned radio lightcurves are shown in Fig.~\ref{fig:timing}C.

\clearpage


\putbib

\clearpage
\section*{Acknowledgments}
We thank the NICER, Swift, XMM-Newton, VLA, ATCA, e-MERLIN, and VLBA teams for approving our ToO/DDT requests. 
The National Radio Astronomy Observatory is a facility of the National Science Foundation operated under cooperative agreement by Associated Universities, Inc.
The Australia Telescope Compact Array is part of the Australia Telescope National Facility (grid.421683.a) which is funded by the Australian Government for operation as a National Facility managed by CSIRO. We acknowledge the Gomeroi people as the traditional owners of the Observatory site. 
e-MERLIN is a National Facility operated by the University of Manchester at Jodrell Bank Observatory on behalf of STFC, part of UK Research and Innovation.

This research used data obtained with the Dark Energy Spectroscopic Instrument (DESI). DESI construction and operations is managed by the Lawrence Berkeley National Laboratory. This material is based upon work supported by the U.S. Department of Energy, Office of Science, Office of High-Energy Physics, under Contract No. DE–AC02–05CH11231, and by the National Energy Research Scientific Computing Center, a DOE Office of Science User Facility under the same contract. Additional support for DESI was provided by the U.S. National Science Foundation (NSF), Division of Astronomical Sciences under Contract No. AST-0950945 to the NSF’s National Optical-Infrared Astronomy Research Laboratory; the Science and Technology Facilities Council of the United Kingdom; the Gordon and Betty Moore Foundation; the Heising-Simons Foundation; the French Alternative Energies and Atomic Energy Commission (CEA); the National Council of Science and Technology of Mexico (CONACYT); the Ministry of Science and Innovation of Spain (MICINN), and by the DESI Member Institutions: www.desi.lbl.gov/collaborating-institutions. The DESI collaboration is honored to be permitted to conduct scientific research on Iolkam Du’ag (Kitt Peak), a mountain with particular significance to the Tohono O’odham Nation. Any opinions, findings, and conclusions or recommendations expressed in this material are those of the author(s) and do not necessarily reflect the views of the U.S. National Science Foundation, the U.S. Department of Energy, or any of the listed funding agencies.

This study used observations collected at the European Organisation for Astronomical Research in the Southern Hemisphere, Chile, as part of ePESSTO+ (the advanced Public ESO Spectroscopic Survey for Transient Objects Survey). ePESSTO+ observations were obtained under ESO program ID 112.25JQ.

We thank Pu Du for coordinating the Lijiang 2.4m Telescope observations, which provided two spectra taken on 2024-02-24.  
We also thank Ning Jiang, Wenda Zhang, Youjun Lu, Song Wang, and Jianmin Wang for the helpful discussions, and Wu Jiang for his help with the VLBA data analysis.

\paragraph*{Funding:}

This research was supported by the National Natural Science Foundation of China (NSFC) under grant numbers 12588202, 12173103, 12261141691, 12473012 and 12203041; by the New Cornerstone Science Foundation through the New Cornerstone Investigator Program and the XPLORER PRIZE; by the Strategic Priority Program of the Chinese Academy of Sciences under grant numbers XDB41000000 and XDB0550203; by ANID, Millennium Science Initiative, ICN12\_009; and by a research grant (VIL60862) from VILLUM FONDEN.
SdP acknowledges support from ERC Advanced Grant 789410.
P.C. acknowledges support via Research Council of Finland (grant 340613).
SNZ acknowledges support from the National Natural Science Foundation of China (grant Nos. 12333007).
MN is supported by the European Research Council (ERC) under the European Union’s Horizon 2020 research and innovation programme (grant agreement No.~948381) and by UK Space Agency Grant No.~ST/Y000692/1.
TP acknowledges the financial support from the Slovenian Research Agency (grants I0-0033, P1-0031, J1-8136, J1-2460 and Z1-1853).
H.Z. acknowledges National Key R\&D Program of China (grant Nos. 2023YFA1607804, 2022YFA1602902, and 2023YFA1608100), and National Natural Science Foundation of China (NSFC; grant Nos. 12120101003, 12373010, and 12233008).
L.G. acknowledges financial support from AGAUR, CSIC, MCIN and AEI 10.13039/501100011033 under projects PID2023-151307NB-I00, PIE 20215AT016, CEX2020-001058-M, ILINK23001, COOPB2304, and 2021-SGR-01270.
CPG acknowledges financial support from the Secretary of Universities and Research (Government of Catalonia) and by the Horizon 2020 Research and Innovation Programme of the European Union under the Marie Sk\l{}odowska-Curie and the Beatriu de Pin\'os 2021 BP 00168 programme, from the Spanish Ministerio de Ciencia e Innovaci\'on (MCIN) and the Agencia Estatal de Investigaci\'on (AEI) 10.13039/501100011033 under the PID2023-151307NB-I00 SNNEXT project, from Centro Superior de Investigaciones Cient\'ificas (CSIC) under the PIE project 20215AT016 and the program Unidad de Excelencia Mar\'ia de Maeztu CEX2020-001058-M, and from the Departament de Recerca i Universitats de la Generalitat de Catalunya through the 2021-SGR-01270 grant. MPT acknowledges financial support from the Severo Ochoa grant CEX2021-001131-S and from the Spanish grant PID2023-147883NB-C21, funded by MCIU/AEI/10.13039/501100011033, as well as support through ERDF/EU.
T.-W.C. acknowledges the Yushan Fellow Program by the Ministry of Education, Taiwan for the financial support (MOE-111-YSFMS-0008-001-P1). FGX and LHW were supported in part by National SKA Program of China (grant Nos. 2020SKA0110100 and 2020SKA0110200), and by National Natural Science Foundation of China (NSFC; grant Nos. 12373017, 12192220 and 12192223). FO acknowledges support from MIUR, PRIN 2020 (grant 2020KB33TP) ``Multimessenger astronomy in the Einstein Telescope Era (METE)'' and from INAF-MINIGRANT (2023): ``SeaTiDE - Searching for Tidal Disruption Events with ZTF: the Tidal Disruption Event population in the era of wide field surveys''.

\paragraph*{Author contributions:}
Conceptualization: Y.N.W., W.H.L., S.N.Z., J.F.L., S.D.P., R.D.B., A.Y.Y., X.W.C., E.L.Q., X.W.S., F.O., and D.A.
Methodology: Y.N.W., Z.K.L., W.H.L., S.N.Z., L.J., J.F.L., B.Z., F.G.X., D.F.B., D.R.P., S.D.P., A.Y.Y., and D.A.
Software: Y.N.W., Z.K.L., L.H.W., W.H.L., S.D.P., Y.H., and A.Y.Y.
Validation: Y.N.W., Z.K.L., L.H.W., W.H.L., Y.H., A.Y.Y., P.C., Y.W.Z., W.X.L., H.Z.W., T.E.M., and J.Y.W.
Formal analysis: Y.N.W., Z.K.L., L.H.W., W.H.L., L.J., F.G.X., Y.H., A.Y.Y., R.R.C., Y.W.Z., X.W.C., H.Z.W., D.W., and Z.H.Y.
Investigation: Y.N.W., Z.K.L., W.H.L., S.D.P., R.D.B., L.Y., Z.H.Y., F.G.X., D.F.B., X.W.C., H.F., Y.H., A.Y.Y., A.L.W., P.C., D.W., W.X.L., M.N., G.L., C.P.G., T.P., T.W., J.Y.W., W.J.G., L.G., and T.W.C.
Resources: Y.N.W., L.H.W., Y.H., A.Y.Y., J.W., H.Z., W.J.G., G.L., J.P.A., T.E.M., Y.L.Q., C.I., T.P., L.G., and T.W.C.
Data curation: Y.N.W., Z.K.L., L.H.W., S.Y.W., A.Y.Y., L.Y., A.L.W., Y.W.Z., Z.H.Y., J.W., G.L., W.J.G., C.I., J.P.A., M.G., and T.P.
Writing—original draft: Y.N.W., Z.K.L., L.H.W., W.H.L., L.J., J.F.L., S.D.P., R.D.B., D.F.B., Z.H.Y., and H.Z.
Writing—reviewing and editing: Y.N.W., Z.K.L., L.H.W., W.H.L., L.J., S.N.Z., J.F.L., F.G.X., D.F.B., H.F., S.D.P., X.W.S., R.D.B., Y.H., A.Y.Y., P.C., D.W., G.L., J.P.A., M.N., L.M., T.E.M., C.P.G., T.W., M.G., F.O., D.A., L.G., M.P.T., and T.W.C.
Visualization: Y.N.W., Z.K.L., L.H.W., W.H.L., S.D.P., R.D.B., J.Y.W., and P.C.
Supervision: Y.N.W., W.H.L., and J.F.L.
Project administration: Y.N.W., J.F.L., and J.P.A.
Funding acquisition: Y.N.W., W.H.L., J.F.L., and C.I.

\paragraph*{Competing interests:}
The authors declare that there are no competing interests. 

\begin{figure}[htbp!]
\centering
\includegraphics[width=1\textwidth]{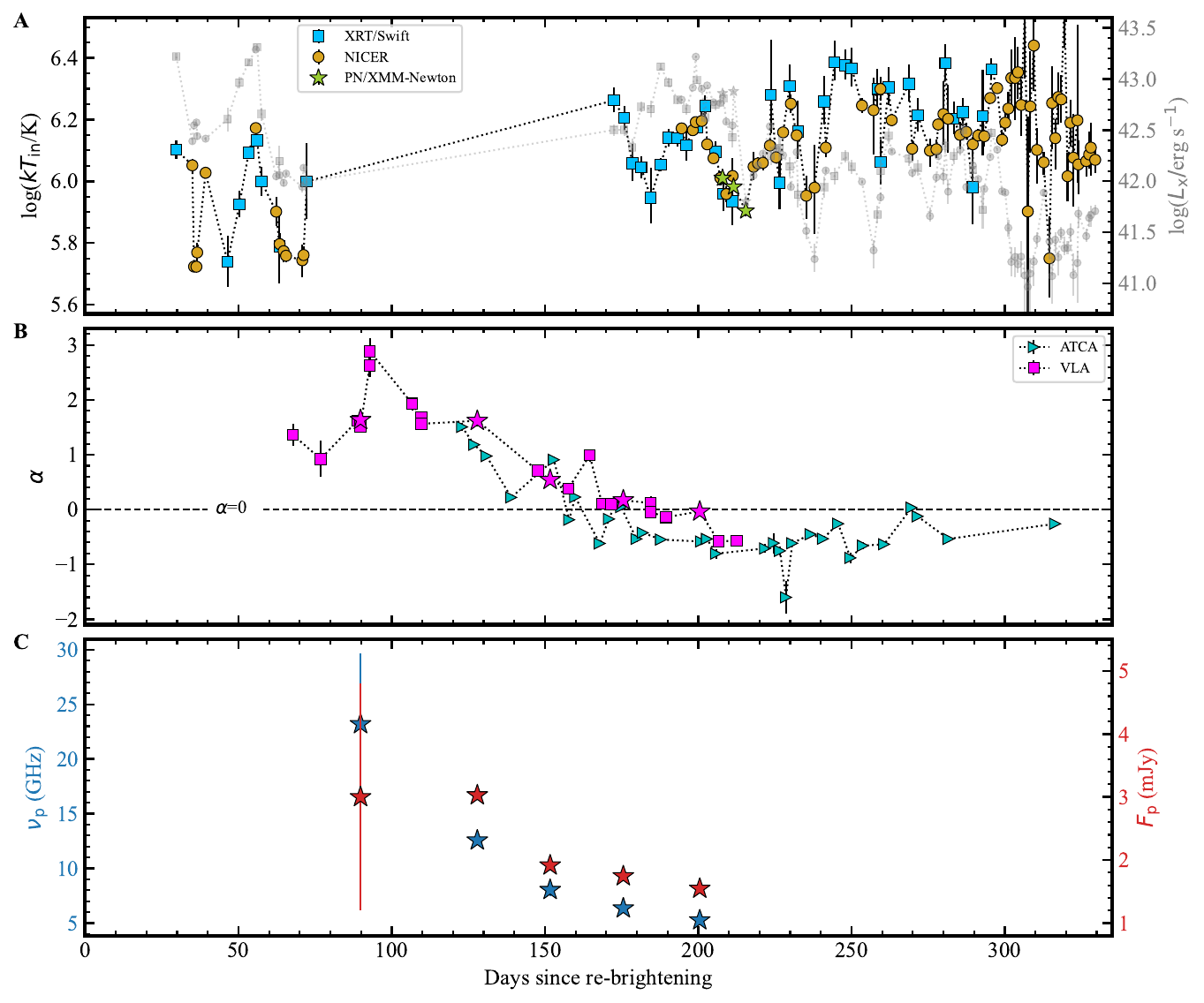}
\caption{\small \textbf{Temporal evolution of spectral parameters.}
(\textbf{A}) The \texttt{diskbb} temperature derived from X-ray spectra. The luminosity evolution is depicted with grey symbols to illustrate its correlation with temperature; the two parameters generally co-evolve, with the temperature reaching its peak alongside luminosity during the first 300 days. (\textbf{B}) The in-band spectral index ($F \propto \nu^\alpha$) derived from the VLA (4--8~GHz) and ATCA (4--10~GHz) data.  The stars indicate the VLA observations used for the radio SED modeling (Supplementary~\ref{sec:radio_sed}). (\textbf{C}) The peak frequency, $\nu_{\rm p}$, and the peak flux, $F_{\rm p}$, derived from the SED modeling.}\label{fig:spec}
\end{figure}

\begin{figure}[htbp!]
\centering
\includegraphics[width=1\textwidth]{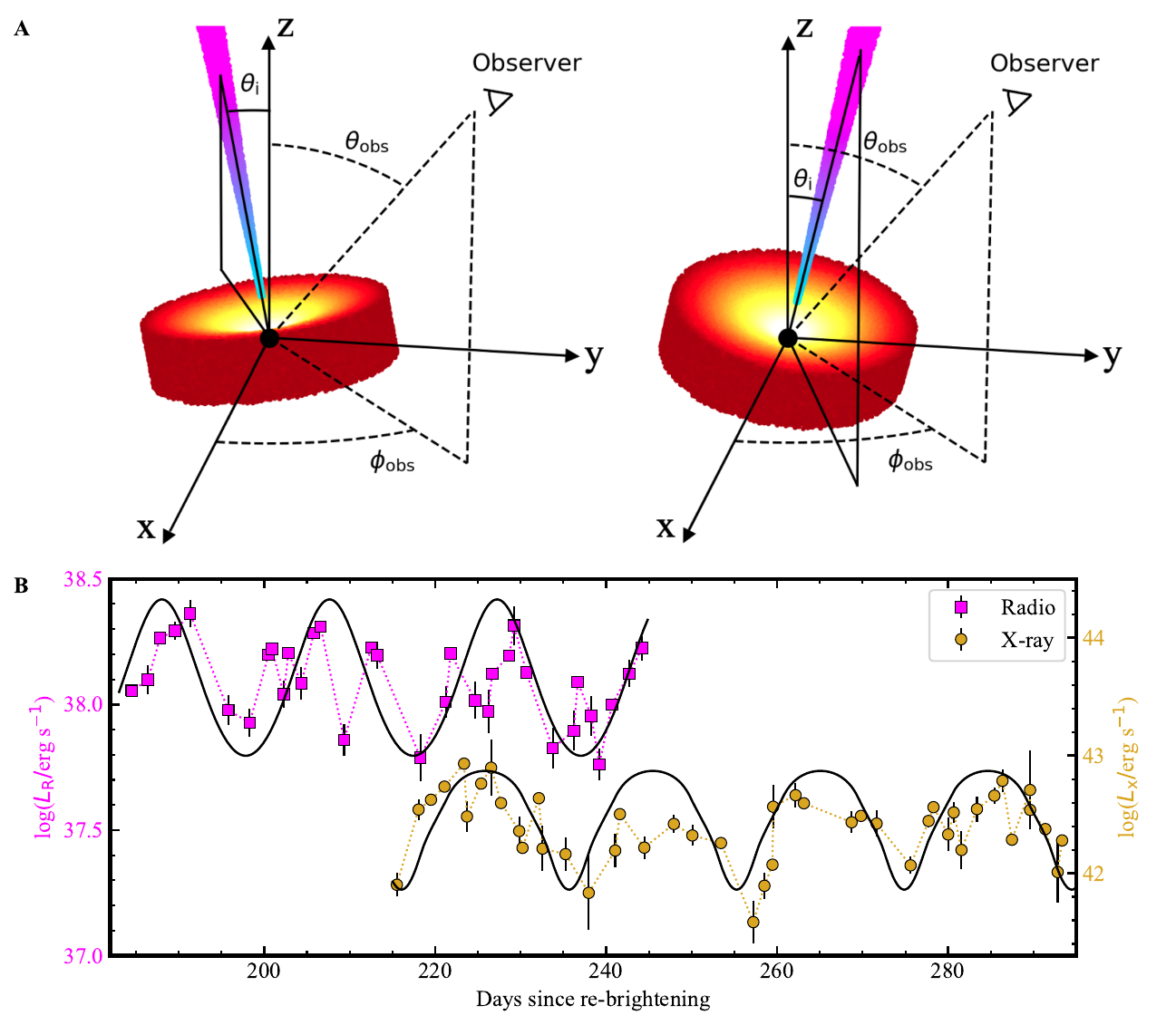} 
\caption{\small \textbf{The disk-jet precession model.}
(\textbf{A}) Schematics of the proposed disk-jet precession model. $\theta_{\rm obs}$ and $\theta_{\rm i}$ represent the viewing angle of the system and the disk/jet precession angle around the black hole axis, respectively. The left and right plots correspond to the phases of the X-ray and radio variations when the luminosity is relatively low and high, respectively.
(\textbf{B}) Comparison of the disk-jet precession model (the lower and upper black curves) with X-ray (0.1--2 keV) and radio (5--6~GHz) observations. In the presented model, we adopted a BH mass of $M_\bullet=10^{6.7} M_{\odot}$, a scaleheight ratio of $H/R\sim1$, and an outer disk radius equal to the circularization radius of the debris. 
As a result, we determined an inclination angle of $\theta_{\rm obs}\sim37.8-38.9^\circ$, a disk/jet precession angle of $\theta_{\rm i} \sim 14-15^\circ$, and a Doppler factor of $\Gamma\sim1.2-1.6$.}\label{fig:schematics_model}
\end{figure}

\paragraph*{Data and materials availability:}
The data used to generate Figs.~\ref{fig:lc}, \ref{fig:spec}, \ref{fig:timing}, and \ref{fig:schematics_model}, along with the Python codes for the Lomb-Scargle periodogram and cross-correlation function, are available on Zenodo ({\href{ https://doi.org/10.5281/zenodo.17649346}{https://doi.org/10.5281/zenodo.17649346}}).

\clearpage


\subsection*{Supplementary materials}
Sections S1 to S6\\
Figures S1 to S5\\

\clearpage 
\vfill\eject

\begin{figure}[htbp!]
\centering
\includegraphics[width=0.85\textwidth]{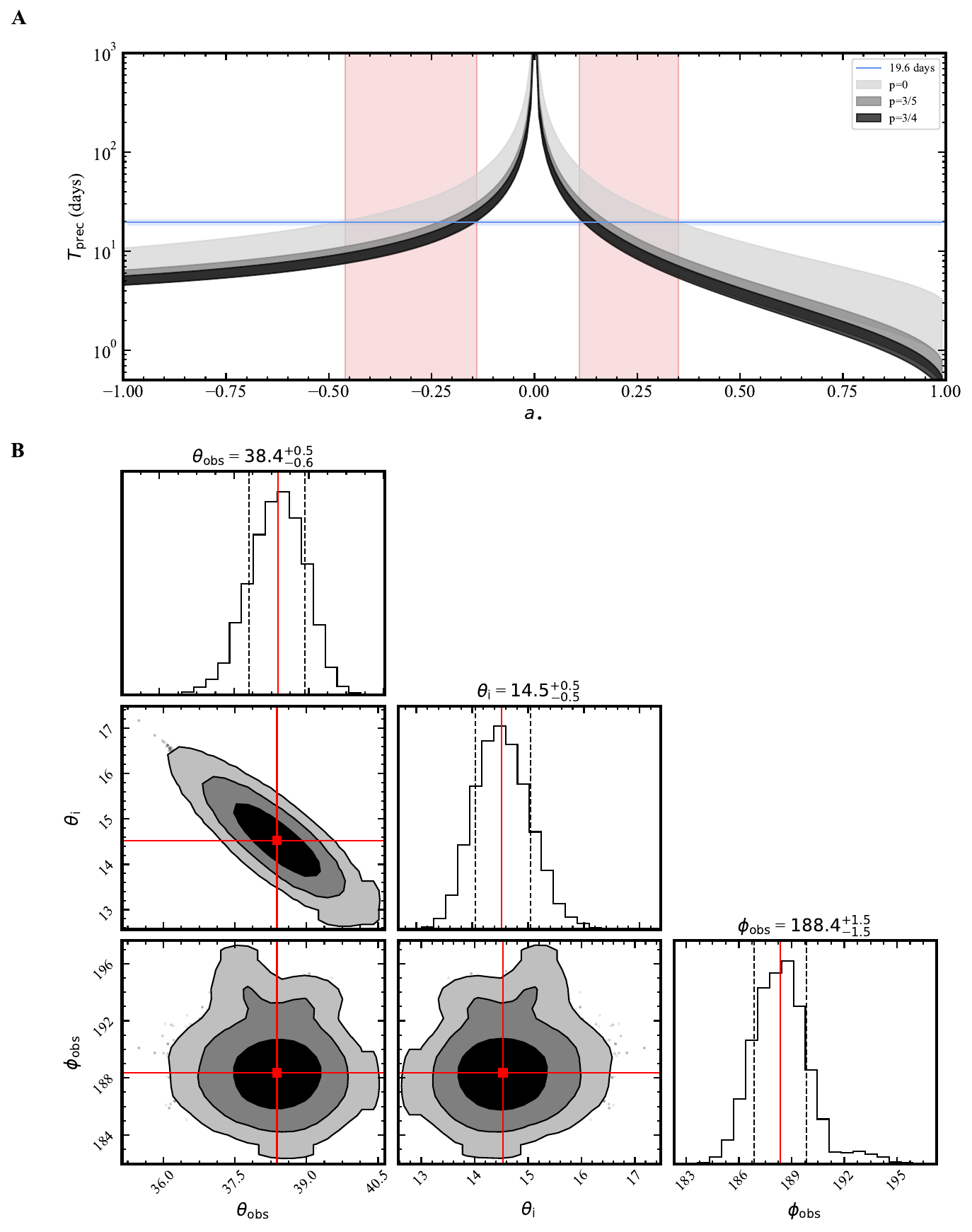}
\caption{\small \textbf{Estimation of system parameters.} (\textbf{A}) Disk precession period vs BH spin when adopting various BH masses and surface density profile of the accretion disk. The gray and the blue shades correspond to a BH mass of $10^{6.7\pm0.5}~M_{\odot}$ and a period of $19.6\pm1.5$~days. The overlap of the two shades constrains the BH spin to ranges of $-0.46 < a_{\bullet} < -0.14$ and $0.11 < a_{\bullet} < 0.35$. (\textbf{B}) Contour plots showing the best-fitting parameters for our proposed disk precession model. }\label{fig:MCMC_precession}
\end{figure}

\begin{table}[htbp!]
\centering
\caption{Summary of the spectroscopic observations}
\begin{tabular}{ccccc}
\hline
Epochs&Telescope/Instrument & Grism & Observation Date (UTC)&Exposure time (s)\\
\hline
\hline
4d &P60/SEDM$^\natural$&- &2024-01-13T06:00:32&$2700$ \\
43d &FLOYDS-S$^\natural$& - &2024-02-13T10:36:20&$1800$ \\
55d &Lijiang 2.4 m/YFOSC& G14@$2.5^{\prime\prime}$&2024-02-24T12:16:06&$1800$\\
55d &Lijiang 2.4 m/YFOSC& G8@$2.5^{\prime\prime}$&2024-02-24T12:51:50&$1800$\\
61d &Xinglong 2.16 m/BFOSC&G4@$1.8^{\prime\prime}$&2024-03-02T11:32:36&$3\times1800$\\
65d &Xinglong 2.16 m/BFOSC&G4@$1.8^{\prime\prime}$&2024-03-06T11:32:32&$2\times1800$\\
73d &NTT/EFOSC2&G11@$1.0^{\prime\prime}$ &2024-03-14T00:20:06&$1200$ \\
239d &NTT/EFOSC2&G11@$1.0^{\prime\prime}$ &2024-08-27T09:11:59&$1200$ \\
\hline 
\hline
\end{tabular}\label{tab:opt_sum}
    \vspace {5pt} 
    \begin {minipage}{\textwidth}
    {\small Note. $^\natural$Spectra from P60 and FLOYDS were downloaded directly from the Transient Name Server (https://www.wis-tns.org/object/2020afhd). The epochs correspond to the times offset from MJD 60310. Spectra with a SNR below 10 are excluded from the table, as they are not utilized in the subsequent analysis.
    }
    \end {minipage}
\end{table}

\newpage


\renewcommand{\thefigure}{S\arabic{figure}}
\renewcommand{\thetable}{S\arabic{table}}
\renewcommand{\theequation}{S\arabic{equation}}
\renewcommand{\thepage}{S\arabic{page}}
\setcounter{figure}{0}
\setcounter{table}{0}
\setcounter{equation}{0}
\setcounter{page}{1} 


\begin{center}
\section*{Supplementary Materials for\\ Detection of disk-jet co-precession in a tidal disruption event}

Yanan Wang et al.\\
\small Corresponding authors. Email: wangyn@bao.ac.cn, leiwh@hust.edu.cn, huangyang@ucas.ac.cn, and jfliu@nao.cas.cn\\
\end{center}

\subsubsection*{This PDF file includes:}
Sections S1 to S6\\
Figures S1 to S5\\



\newpage


%
%
\newpage
\setcounter{page}{1}
\renewcommand{\theequation}{S\arabic{equation}}
\section*{{\Huge Supplementary Materials}}
\section{Observations and Data Analysis} \label{sec:data}


\subsection{UV/optical data}
\subsubsection{Ultraviolet Optical Telescope (UVOT)/Neil Gehrels Swift Observatory (Swift)}
\noindent Swift \cite{Roming2005} performed 66 target-of-opportunity (ToO) observations of AT2020afhd from January 31 to November 22, 2024 (PI: Y. Wang, E. Hammerstein, N. Jiang, and S. Huang). When multiple snapshots were included in the observations of each individual filter, the \textsc{uvotimsum} task was used to combine all exposures.
We used an 8$^{\prime\prime}$ aperture centered on the target position as the source, and a nearby 40$^{\prime\prime}$ source-free region as the background.
The target was observed with all or some of the six filters, i.e., UVW2 (central wavelength, 1928 \AA), UVM2 (2246 \AA), UVW1 (2600 \AA), U (3465 \AA), B (4392 \AA) and V (5468 \AA), across different observations. The UVOT magnitudes have been corrected for Galactic extinction and host galaxy contributions (see Supplementary~\ref{sec:host_galaxy}).


\subsubsection{Asteroid Terrestrial-impact Last Alert System (ATLAS)}
\noindent The ATLAS telescopes conducted observations by scanning the visible sky with a cadence of one to two nights, using broadband filters as part of the standard asteroid search program \cite{Heinze2018, Tonry2018}. Specifically, two filters were employed: the c-band (4200--6500~\AA) and o-band (5600--8200~\AA). We retrieved difference photometry data of the source from the ATLAS forced photometry server \cite{Shingles2021}, selecting observations between MJD~60300 and 60640 with a signal-to-noise ratio (SNR) $>$ 5. To enhance the accuracy of flux measurements, intra-night observations were combined into several epochs using a weighted mean, with weights assigned as the inverse variances $1/\sigma^2$ of each individual observation. The uncertainty for each epoch was calculated by the following formula: 
$$ \sigma_{\rm epoch}^2 = \frac{1}{\sum_{i=1}^N(\frac{1}{\sigma_i^2})} + \frac{1}{N-1}\sum_{i=1}^N(m_i - m_{\rm epoch})^2.$$
We applied iterative 3-sigma clipping within each epoch to remove outliers from intra-night observations. Due to its superior sampling, only the o-band data were included in this work. Finally, we selected epochs with $\frac{m_{\rm epoch}}{\sigma_{\rm epoch}} > 3$ and corrected those magnitudes for Galactic extinction.

\subsubsection{Visible Telescope (VT)/Space-based multi-band astronomical Variable Objects Monitor (SVOM)}
We observed \abw using the VT onboard the SVOM mission \cite{Wei2016}. VT utilizes a dichroic beam splitter to separate light into two channels, enabling simultaneous observations in two bands: VT-$B$ (4000--6500~{\AA}) and VT-$R$ (6500--10000~{\AA}).
The observations were conducted from November 3 to 20, 2024, spanning approximately 20 days with 11 orbits. During each orbit, $\sim100$ images were obtained in both bands, each with an exposure time of 20 seconds.

Data reduction followed standard Image Reduction and Analysis Facility (IRAF, \cite{Tody1986,Tody1993}) processes, including bias, dark, and flat-field corrections. Images from each orbit were combined to improve the SNR, and aperture photometry was performed on the combined images. A 7-pixel radius aperture was used for the target, with an annulus of 10--15 pixels defining the background. Aperture corrections were applied to derive source magnitudes.
For accurate measurements, differential photometry was performed using four bright reference stars in the same field. The average flux of these reference stars was used to calculate the differential magnitude of the target. Magnitude uncertainties account for both measurement and systematic errors, and all magnitudes were calibrated to the AB magnitude system. 

\subsubsection{New Technology Telescope (NTT)}
\noindent NTT is a 3.58m telescope located at the European Southern Observatory (ESO) La Silla Observatory in Chile. Two optical spectra were obtained on March 14 and August 27, 2024, via ESO as part of the ePESSTO+ program, using the ESO Faint Object Spectrograph and Camera (EFOSC2; \cite{Buzzoni1984}). Both spectra were taken with grism11 (covering 3380--7520 \AA, with a resolution of approximately 15.8 \AA) using a single 1200 s exposure. Data reduction was performed using version 3.0.1 of the PESSTO pipeline \cite{Smartt2015}.\\

\subsubsection{Lijiang 2.4-m Telescope}
Lijiang 2.4-m Telescope is located in Lijiang of the Yunnan Observatories. Two optical spectra were obtained on February 14, 2024, using G14 (covering 3600--7460~\AA) and G8 (covering 5100--9600~\AA), each with a resolution of approximately 8.3~\AA~and a single 1800~s exposure. Data reduction was carried out using IRAF.

\subsubsection{Xinglong 2.16-m Telescope}
Optical spectra of AT2020afhd were extensively collected using BFOSC with a new Grism G4, mounted on the 2.16-meter telescope at the Xinglong Observatory in Hebei Province, operated by the National Astronomical Observatories of the Chinese Academy of Sciences.
Between February 9 and March 20, 2024, a total of 11 optical spectra covering the wavelength range of 3850--7500~\AA~ were obtained. The typical exposure times ranged from 1800 to 2400 seconds, depending on the specific weather conditions.
Most spectra were observed using a slit width of $1.8^{\prime\prime}$, corresponding to a resolution of approximately $~20$~\AA. However, a few spectra were taken with a slit width of $2.3^{\prime\prime}$, yielding a resolution of $25$~\AA, due to poor seeing conditions.
After bias subtraction and flat-field correction, the spectra were reduced using a specialized Python package specifically designed for the BFOSC instrument on the 2.16-m telescope.

\subsubsection{Dark Energy Spectroscopic Instrument (DESI)}
DESI is amounted on the 4-meter Mayall Telescope, located at the Kitt Peak National Observatory in Arizona, USA \cite{DESI2016a,DESI2022b}. 
On February 19, 2022, DESI successfully obtained one spectrum of \abw. This spectrum was selected from the DESI Bright Galaxy Survey \cite{Hahn2023}, with target sources coming from the DESI Legacy Imaging Survey \cite{Dey2019,Zou2017}. The spectral range spans from 3600~\AA~ to 9800~\AA, with an exposure time of $780$ s and a spectral resolution of $R$~=~2000--5000. The observed data were processed using the DESI spectroscopic data processing pipeline \cite{Guy2023}.

\subsection{X-ray data}

\subsubsection{X-Ray Telescope (XRT)/Swift}

\noindent XRT observations were performed in photon-counting mode, with a total exposure time of $\sim132.6$ ks. The XRT spectra were downloaded directly from the UK Swift Science Data Centre \cite{Evans2007,Evans2009}. To constrain the continuum, only observations with more than 8 counts were included in the analysis. The spectra were grouped using the FTOOL {\sc ftgrouppha} with \texttt{grouptype=optmin} and \texttt{groupscale=3}. Consequently, Cash statistics \cite{Cash1979} was applied for spectral fitting. As the source is background-dominated above 2 keV, the XRT spectra were fitted in the 0.3--2 keV energy band only.

\subsubsection{Neutron Star Interior Composition Explorer (NICER)}
\noindent NICER observed AT2020afhd from February 4, to November 24, 2024, with a total exposure time of 156.9 ks (PI: Y. Wang, Y. Yao, and T. Wevers). Raw data were obtained from the HEASARC public archive and processed using the NICER Data Analysis Software (version v2023-08-22\_V011a). Calibration (version xti20221001) and screening of the raw data were performed using {\sc nicerl2} with thresh\_range=``--3.0--38''. Good time intervals (GTIs) with count rates exceeding 1 count per second in the 8--12 keV band were excluded to prevent contamination from background flares. 
For observations with a single GTI and a gap of less than 0.1 days before the next observation, we merged the two using {\sc niobsmerge} to improve the SNR.
Spectra were generated with {\sc nicerl3-spect}, and background contributions were estimated using the SCORPEON model (\href{https://heasarc.gsfc.nasa.gov/docs/nicer/analysis_threads/scorpeon-overview/}{https://heasarc.gsfc.nasa.gov/docs/nicer/analysis\_threads/scorpeon-overview/}). 
The spectra were binned with {\sc ftgrouppha}, using \texttt{grouptype=optmin} and \texttt{groupscale=3}, ensuring a minimum of 3 counts per bin. 
For spectral fitting, the energy range of 0.3--15 keV was used, and the \texttt{con\_norm} parameter was allowed to vary to accurately estimate the background components.

\subsubsection{PN/XMM-Newton}
\noindent XMM-Newton conducted four observations of AT2020afhd (obsIDs: 0932391901, 0953010101, 0953010201, and 0953010301), with a total exposure of 106.5 ks. Raw data were downloaded from the XMM-Newton Science Archive and processed using XMM-Newton’s Science Analysis Software (version 20.0.0). Calibrated and concatenated event lists were generated with the {\sc epproc} task. A source region with a radius of 60$^{\prime\prime}$ was centered on the source using the {\sc centroid} tool in {\sc DS9}. A background region of the same radius was selected in a corner of the same detector chip as the source. To reduce the effects of background flares, we excluded GTIs with a count rate $>$0.4 cts/s in the 10--12 keV band. For obsID 0932391901, the entire lightcurve in the 10--12 keV band exceeded 0.4 cts/s, so this observation was excluded from further analysis. Source and background spectra were extracted using the {\sc evselect} task with PATTERN$\leq$4 and FLAG=0. The corresponding response matrices were generated with the {\sc rmfgen} and {\sc arfgen} tasks. The spectra were then binned using the {\sc specgroup} task, with mincounts=3 and oversample=3. 



\subsubsection{Extended ROentgen Survey with an Imaging Telescope Array (eROSITA)/Spectrum-Roentgen-Gamma (SRG)}
We used eROSITA \cite{Predehl2021} all-sky survey (eRASS1) data, where our source is located in the Skytile 050093 (\href{https://erosita.mpe.mpg.de/dr1/erodat/skyview/skytile/050093/?goto=48.39944,-2.15094}{https://erosita.mpe.mpg.de/dr1/erodat/skyview/skytile/050093/?goto=48.39944,-2.15094}). This region was observed between February 3 and 16, 2020, nearly four years before the TDE.
We analyzed the event file using the eROSITA Science Analysis Software System (eSASS) following the official recommended cookbook \\(\href{https://erosita.mpe.mpg.de/dr1/eSASS4DR1/eSASS4DR1_cookbook/}{https://erosita.mpe.mpg.de/dr1/eSASS4DR1/eSASS4DR1\_cookbook/}). Our source can be clearly identified, and we extracted its spectrum using the {\tt srctool} task. The spectrum was re-binned using the {\sc grppha} task, with mincounts=3. 

\subsection{Radio data}
\subsubsection{Very Large Array (VLA)}
\noindent The VLA conducted 25 observations of AT2020afhd between March 8 and July 31, 2024, as part of the DDT projects 24A-442, 24A-466, and 24A-485 (PI: Y. Wang). Of these, 23 observations were carried out at C band (4--8 GHz) to monitor the short-term evolution of the luminosities. Additionally, 2 observations were conducted across multiple bands to capture the spectral evolution, specifically covering S, C, and Ku bands (4--18 GHz) and L, S, C, and X bands (1--8 GHz), respectively.

The radio data were reduced using the Common Astronomy Software Application package (CASA 6.5.4, \cite{McMullin2007, CASA2022}), following standard data reduction procedures. The calibrated measurement sets were downloaded from the NRAO archive (\href{https://data.nrao.edu/portal/}{https://data.nrao.edu/portal/}), which were calibrated using the automated VLA calibration pipeline (\href{https://science.nrao.edu/facilities/vla/data-processing/pipeline}{https://science.nrao.edu/facilities/vla/data-processing/pipeline}) available in the CASA. 
Flux density and bandpass calibration were conducted using 3C48, 3C147, 3C286, or 3C138, while J0339--0146 was used as a gain calibrator. 
The observation taken on MJD 60395.8, which failed to pass the pipeline's quality assessment, was excluded from the analysis.
Images of the AT2020afhd field were produced using the CASA task \textsc{tclean} with a Briggs robust parameter of 0.5 \cite{Briggs1995}. We measured the flux density of AT2020afhd by fitting a point-source Gaussian model in the image plane with the CASA task \textsc{imfit} and determined the off-source rms using the CASA task \textsc{imstat}. The flux density error was estimated as the quadratic sum of the off-source rms, the fitting error from \textsc{imfit}, and a typical systematic error of 5\% of the source flux density. 
To construct the lightcurve, all C-band observations, including 2 multi-band epochs, were imaged using a reference frequency at 5.5 GHz. The in-band spectral index at C band (VLA-$\alpha$) was calculated by imaging each spectral window (spw) separately for each epoch and fitting a power-law model with Python’s \texttt{curve\_fit} package. 
To capture the spectral evolution, the broadband spectrum at the two multi-band epochs (i.e., observed on MJD 60437.9 and 60510.5) and three C-band epochs (i.e., observed on MJD 60399.8, 60461.7 and 60485.6) were constructed (see the selection criteria in Supplementary~\ref{sec:radio_sed}). Each set of 8 spws was imaged separately for the two multi-band epochs, and each set of 4 spws was imaged separately for the three selected C-band epochs (see fig.~\ref{fig:radio_sed}).

\subsubsection{Australia Telescope Compact Array (ATCA)}
\noindent The ATCA conducted 32 observations of AT2020afhd between May 2 and November 12, 2024, as part of the DDT projects CX567 and CX591, and the Non A-Priori Assignable project C3615 (PI: Y. Wang). All the observations were observed at C band (4--10 GHz).

The radio data were reduced in CASA of the same version for the VLA data, following the tutorial
(\href{https://casaguides.nrao.edu/index.php/ATCA\_Tutorials}{https://casaguides.nrao.edu/index.php/ATCA\_Tutorials}) and additionally using the `rflag' and `tfcrop' modes in the CASA task \textsc{flagdata} to flag the radio-frequency interference. In all observations, 1934--638 was used to calibrate the bandpass response, flux-density scale and polarization leakage, while 0336--019 was used to calibrate the time-variable complex gains. We imaged all the observations for the AT2020afhd field at C band with reference frequencies at 5.5 GHz and 9 GHz, each covering a bandwidth of 2048 MHz, using the CASA task \textsc{tclean} with natural weighting. The flux density and corresponding error measurements of AT2020afhd were the same as those of VLA. We calculated the spectral index for ATCA (i.e., ATCA-$\alpha$) by fitting a power-law to the two-point flux measurements at 5.5 GHz and 9 GHz.

\subsubsection{Enhanced Multi-Element Radio Linked Interferometer Network (e-MERLIN)}
\noindent e-MERLIN conducted 23 observations of AT2020afhd between July 5 and November 18, 2024, as part of the Rapid Response Time projects RR17006, RR18001, and RR18002 (PI: Y. Wang). These observations were carried out at C band, covering frequencies from 4.8 to 5.3 GHz. The data were processed using a CASA-based (version 5.8.0) pipeline \cite{Moldon2021}, which performs data averaging, flags for frequency interference, calibrates, and images at a central frequency of $\sim5$~GHz. 1331+3030 (3C286), 1407+2827 (OQ208), 0319+4130 (3C84) and 0315-0151 were used as flux calibrator, bandpass calibrator, bright ptcal calibrator, and gain calibrator, respectively.


We noted that four observations—RR17006\_C\_004\_20240716, RR17006\_C\_006\_20240723, RR18001\_C\_005\_20240827 and RR18002\_C\_001\_20241105—suffered from excessive visibility weights, resulting in suboptimal images produced by the pipeline. To address this, we flagged these problematic visibilities and re-imaged these observations using the CASA task \textsc{tclean} with the same parameters as the pipeline. RR18002\_C\_004\_20241117 was excluded in the analysis, as no significant signal, i.e., greater than three times the off-source rms, was detected at the target position in the final image. This was likely due to the majority ($\sim$85\%) of the data being flagged and the short exposure ($\sim$2.5 hours on target). The flux density and corresponding error measurements of AT2020afhd followed the same methodology as those for the VLA, including a 10\% systematic error as suggested by the e-MERLIN team.


\subsubsection{Very Long Based Array (VLBA)}
We conducted four VLBA observations of AT2020afhd: on 2024-06-21, 2024-09-03, and 2024-11-01 at C band, and on 2024-06-29 at U band, with at least 8 participating antennas (10 for the first epoch). All observations were part of the DDT project BW161 (PI: Y. Wang). The VLBA data were correlated at the Array Operations Center in Socorro, USA. We phase-referenced our observations to the calibrator J0315--0151 with cycle time of 5 min (1 min on calibrator and 4 min on target) for a total of $\sim$1.5~h each epoch for C band (6.2~GHz) and U band (15~GHz). We used the source J0339-0146 as fringe finder and bandpass calibrator. 
We performed standard a-priori gain calibration using the measured gains and system temperatures of each antenna. The data inspection and flagging, full calibration were done within the National Radio Astronomy Observatory Astronomical Image Processing System \cite{Greisen03}. We also corrected for ionosphere effects and source-structure effects of the phase-reference source in each band \cite{Beasley95}. After the calibration, 20\% of the data from the second and third epochs were discarded due to poor quality, resulting in noisier datasets compared to the first and fourth epochs.

All imaging and deconvolution were performed using DIFMAP \cite{Shepherd94}, with a cell size of 0.2 mas for C band and 0.1 mas for U band. We applied both natural and uniform weighting to search for compact and diffuse emission features (see fig.~\ref{fig:vlba}, shown with natural weighting).
The off-source rms was 22, 58, 28, and 17 $\mu$Jy beam$^{-1}$ for the three epochs down to a resolution of $\sim$2 mas and $\sim$5 mas for the third epoch. The sources appeared all unresolved, with deconvolved sizes smaller than the beam width, and were detected above 5$\sigma$ (total flux density $\sim$ 1 mJy) with brightness temperatures of $\sim$10$^{7}$ K.

The observation log for all the radio epochs at C-band are available on Zenodo. The columns in the log include the observation date, time, MJD, radio telescope, telescope configuration, reference frequency of image, bandwidth, major axis of beam, minor axis of beam, position angle (PA) of beam, flux density, flux density error, off-source rms, and SNR (flux density divided by off-source rms), respectively.




\subsubsection{Evaluation of calibration differences among radio arrays}\label{sec:radioflux_offset}
We conducted radio monitoring of \abw using four long-baseline arrays. Although high-amplitude variations were observed in the data from all four arrays individually, the mixed use of different arrays may have introduced additional offsets in the flux densities. Here, we investigated potential calibration offsets among the VLA, ATCA, and e-MERLIN, and evaluated their impact on the observed variations. The sampling of the VLBA observations was too sparse to conduct this test.

To evaluate the flux density offset between VLA and ATCA, we used the only phase calibrator shared by both arrays, J0339--0146 (0336--019). For each observation, we inspected the output CASA log of the \textsc{fluxscale} task to get the fitted spectrum model of J0339--0146 and measured its flux density at 5.5 GHz. The results showed that J0339--0146 remained relatively stable during the studied periods, with mean flux densities of $\mu=1.30$ Jy (VLA) and $1.59$ Jy (ATCA), and standard deviations of $\sigma=0.11$ Jy for both arrays. This yielded a small coefficient of variation (CV), $\sigma/\mu\sim6.9$--8.5\%, which is substantially lower than those observed in \abw, where $\sigma/\mu\sim37.5$--41.7\% and $\sim25.0$--33.3\% were measured before and after detrending the long-term evolution using equation~\ref{eq:S2}, respectively.
The offset between the mean flux densities of the two arrays is approximately 18--22\%.

Similarly, we selected the bandpass calibrator of our e-MERLIN observations, namely J1407+2827 (OQ208), to evaluate the calibration offsets between the VLA and e-MERLIN arrays. During the studied period, we identified five VLA observations of J1407+2827 at C band using B or BnA configurations from the NRAO archive. These observations utilized the same flux calibrator (3C286) with our e-MERLIN observations, and also demonstrated a partial overlap in UV coverage with our e-MERLIN observations. The five VLA epochs were taken on MJD 60455.2, 60466.0, 60484.3, 60554.680, and 60589.6. These observations were calibrated with CASA 6.5.4 following standard data reduction procedures. We found that J1407+2827 showed comparable amplitude within the same frequency range across the overlapping UV coverage for both arrays.
Additionally, we derived the flux density of J1407+2827 at 5.07 GHz for each observation using the same method described earlier. Overall, the mean flux densities were $\mu=1.60$ Jy and $1.49$ Jy, with standard deviations of $\sigma=0.01$ Jy and $0.12$ for VLA and e-MERLIN, respectively. This again yielded a small CV, $\sigma/\mu\sim0.6$--8.1\%.
Thus, despite the larger scatter in flux densities from the e-MERLIN observations, the flux calibrations between the two arrays were found to be marginally consistent.

In conclusion, the low CV values of J0339-0146 and J1407+2827 at the observed frequency confirmed that the radio variations of \abw were not caused by flux calibrators. The offsets in the mean flux densities of the two calibrators, as determined from the three arrays, were up to 22\%. To fully account for these calibration differences, we applied the maximum offset value of 22\% to reduce the ATCA flux densities, and used this correction consistently throughout the analysis of the radio-X-ray cross-correlation function.

\section{X-ray spectral analysis}\label{sec:xspec}
\noindent X-ray spectral analysis was performed using XSPEC version 12.13.1 \cite{Arnaud1996}, with C-stat applied for goodness-of-fit testing. For XRT and NICER, the X-ray spectrum is ultra-soft, predominantly background-dominated above 2 keV, and is well described by an absorbed multi-color blackbody component; while for the PN spectrum, where individual exposures are significantly longer than those of the other two facilities, we extended the fitting to the 0.3--10 keV energy band to assess any hard excess. 

In addition to using \texttt{tbabs\_Gal} to account for Galactic absorption, we introduced another component, \texttt{tbabs\_Loc}, to evaluate the presence of local absorbers from the host galaxy. The value of \texttt{tbabs\_Gal} was fixed at $4.72 \times 10^{20}$ cm$^{-2}$, as determined from the HI4PI survey \cite{HI4PI2016}. However, due to the narrow energy band and relatively low flux, \texttt{tbabs\_Loc} could not be well constrained when fitting each spectrum individually. To robustly determine \texttt{tbabs\_Loc}, we adopted a two-step approach: 1) jointly fitting spectra from the same facility with the data obtained before MJD 60530 when the target was in a high-luminosity phase, and comparing the results to ensure consistency; 2) jointly fitting spectra obtained at high and low luminosities, respectively, to evaluate whether absorption contributes to the observed modulation.
The initial model applied was \texttt{tbabs\_Gal*zashift*tbabs\_Lcol*diskbb} for the XRT and NICER data, and \texttt{tbabs\_Gal*zashift*tbabs\_Lcol*thcomp*diskbb} for the PN data. The redshift was fixed at 0.027. The abundance table and the photoelectric cross-section for \texttt{tbabs} were set to wilm \cite{Wilms2000}, and vern \cite{Verner1996}, respectively. 
The unabsorbed flux of the \texttt{diskbb} component was calculated using the \texttt{cflux} command in the 0.3--2 keV.

We first applied step 1), jointly fitting the spectra with \texttt{tbabs\_Lcol} linked, and obtained values consistent with zero. The 95\% upper limits for \(n_{\rm H\_Loc}\) were \(0.2 \times 10^{20}~\text{cm}^{-2}\) for NICER, \(2.6 \times 10^{20}~\text{cm}^{-2}\) for XRT, and \(0.9 \times 10^{20}~\text{cm}^{-2}\) for PN. 
Next, we applied step 2), jointly fitting the spectra above and below the luminosity of \(9.2 \times 10^{41}~\text{erg~s}^{-1}\) with \texttt{tbabs\_Lcol} linked. Similar to step 1, only upper limits were obtained: \(0.03 \times 10^{20}~\text{cm}^{-2}\) for higher luminosities and \(0.2 \times 10^{20}~\text{cm}^{-2}\) for lower luminosities. If removing \texttt{tbabs\_Lcol} from the fit, the flux measurements across different facilities and epochs were impacted by less than 6\%, within the 1-$\sigma$ uncertainties.
Overall, we concluded that \texttt{tbabs\_Lcol} is negligible and therefore excluded it from our spectral analysis. 

An absorption feature around 0.9 keV was observed in some of the observations from the three facilities spanning from July 13 to 27, suggesting the presence of clumpy outflows along the line of sight. A detailed analysis of these absorption lines is presented in a separate study \cite{Lin2025}.

Additionally, one can see from the model setup, a hard excess above 1 keV was evident in the XMM-Newton observations and could be modeled with a \texttt{thcomp} component convolved with the \texttt{diskbb} component. This suggests that the observed hard excess may result from inverse Comptonization of the disk photons. The \texttt{thcomp} component contributes less than $10\%$ to the total flux.

Unlike the above thermal blackbody spectrum, the SRG spectrum obtained prior to the TDE exhibits a \texttt{powerlaw} profile with a photon index of $2.07\pm0.57$. The corresponding 0.3--2 keV luminosity is $10^{41.8\pm0.1}\rm~erg~s^{-1}$, which exceeds the luminosity measured after day 300 following the re-brightening, when the X-ray emission underwent a second rapid decline. This suggests that the AGN activity observed in 2020 had likely weakened by the time the TDE occurred.

\section{Radio long-term evolution}\label{sec:radio_sed}
\noindent Besides the short-term variability, the radio emission also presents a long-term evolution, gradually increasing since its first detection, peaking around 168 days after the re-brightening, and then decaying throughout the studied period. 
To understand this, we calculated the in-band spectral index $\alpha$ (defined as $F \propto \nu^\alpha$) using the C-band VLA (4--8 GHz) and ATCA (4--10 GHz) data and analyzed the broadband spectral energy distributions (SEDs). 
As shown in Fig.~\ref{fig:spec}B, VLA-$\alpha$ and ATCA-$\alpha$ show similar evolution with the latter remaining slightly more negative. Compared to VLA-$\alpha$, ATCA-$\alpha$ was calculated over a broader and higher frequency range.
Interestingly, the transition of ATCA-$\alpha$ from positive to negative coincides with the shift of the luminosity at 5--6~GHz from increasing to decreasing. 
This overall long-term trend in both the lightcurve and spectral index suggests that the source transitioned from being optically thick to optically thin at C band, likely due to an expanding emitting region. 
In the following, we examined this possibility with the broadband SED fitting.

The typical characteristic of an optically-thin synchrotron spectrum is a decreasing flux density with frequency, that is, $F_\nu \propto \nu^\alpha$ with $\alpha < 0$. However, the synchrotron spectrum from very compact and bright sources such as TDEs can be optically thick at low frequencies. We can introduce a parameter $\nu_\mathrm{a}$ that marks this transition from an optically-thin ($\alpha<0$) SED at $\nu \gg \nu_\mathrm{a}$ to a self-absorbed SED at $\nu \ll \nu_\mathrm{a}$ ($\alpha=2.5$). Then, following e.g.~\cite{Wang2023}, the radio SED of the TDE can be parameterized as:
\begin{equation} 
\label{eq:radio_SED}
F_\nu^\mathrm{TDE} = F_0 \left( \frac{\nu}{\nu_0} \right)^{5/2} \left[ 1 + \left(\frac{\nu}{\nu_\mathrm{a}}\right)^{s_2 (b_2 - b_3)} \right]^{-1/s_2},
\end{equation}
where $F_0$ is the flux density at a reference frequency $\nu_0$ (which we fix at 5~GHz), $b_2 = 5/2$, $b3 = (1 - p)/2$, with $p$ being the spectral index of the relativistic electron population ($N(\gamma_\mathrm{e}) \propto \gamma_\mathrm{e}^{-p}$, where $\gamma_\mathrm{e}$ the electron Lorentz factor). The parameter $s_2$ is given by $s_2 = 1.47 - 0.21 p$ \cite{Granot2002,Cendes2024}. We note that relativistic electrons emitting at $\sim$GHz frequencies evolve on very long timescales, and it is thus reasonable to assume that $p$ remains constant along all the observations.

Furthermore, our observations at $\nu \lesssim 2$~GHz suggest an additional component in the SED, probably related with free--free or diffuse synchrotron emission from the host galaxy. Given the non-detection of any radio emission at 2--4~GHz by the VLASS survey previous to the TDE, we can assume that the galactic emission has a flat or inverted spectrum. For simplicity and to avoid introducing additional parameters in the model, we parameterize this emission as $F_\nu^\mathrm{host} = F_1 \left(\frac{\nu}{\nu_\mathrm{0}}\right)^{-0.1}$, with the spectral index of $-$0.1 being typical of free--free emission. We assume that this emission is steady, and that the total emission is simply $F_\nu(t) = F_\nu^\mathrm{host} + F_\nu^\mathrm{TDE}(t)$.

We therefore have $F_0$, $\nu_\mathrm{a}$, $p$, and $F_1$ as free parameters. We use the Python package \texttt{Bilby} \cite{bilby} for fitting the SEDs of each epoch independently using the standard sampler for Markov Chain Monte Carlo \texttt{emcee} \cite{emcee}. In general, we adopt flat priors over a wide range of parameters. However, for some parameters that are poorly constrained in some epochs we instead set gaussian priors in an iterative approach. First, we fit the epoch in MJD~60437.9 for which we have a good frequency coverage up to $\lesssim 2$~GHz, and obtain a well-constrained value for $\log{F_1}$. Second, we use that value and its uncertainty to define a Gaussian prior for $\log{F_1}$ in MJD~60510.5, which is the only epoch for which we count with data at 10--20 GHz, allowing us to fit $p=2.32\pm0.17$. Next, we use this value of $p$ as a Gaussian prior in the previous epoch of MJD~60437.9 to refine the value of $\log{F_1}$. 

The radio lightcurve has been modulated by both the long-term and the short-term variability. The long-term evolution seems to be able to reveal the intrinsic emission from the radio counterpart dominated by the mass accretion and ejection efficiency. Therefore, to fairly monitor the SED evolution, we first detrend the radio lightcurve using a broken power-law function:
\begin{equation} 
f(x) = 
\begin{cases} 
b \left( \frac{x}{x_\mathrm{break}} \right)^{a_1}, & \text{if } x < x_{\rm break} \\
b \left( \frac{x}{x_\mathrm{break}} \right)^{a_2}, & \text{if } x \ge x_{\rm break}
\end{cases}\label{eq:S2}
\end{equation} 
where the best-fitting break time is $x_{\rm break} = 167.9 \pm 6.3$ day. The slope during the rising phase is $a_1 = 1.8 \pm 0.4$, and the slope during the decaying phase is $a_2 = -1.2 \pm 0.2$. The normalization parameter is $b = 1.8 \pm 0.1 \times 10^{38}$\,$\mathrm{erg~s^{-1}}$. We then select three additional epochs that show the least deviation from the broken power-law function and fitted the epochs, for which only C-band data is available, using the obtained Gaussian priors for $\log{F_1}$ and $p$. The results are shown in fig.~\ref{fig:radio_sed}.

Next, we derive the position of the peaks of all SEDs and their errorbars, $\nu_\mathrm{p}$ and $F_\mathrm{p}$, by sampling the posteriors of each fit. For the five VLA epochs fitted in fig.~\ref{fig:radio_sed}, we obtain that $\nu_\mathrm{p}$ decreases monotonically from $23.2\pm 6.3$ GHz to $5.26\pm 0.18$ GHz, as expected in an expanding source. Similarly, $F_\mathrm{p}$ also decreases from $\sim$3~mJy to $\approx$1.5~mJy (see Fig.~\ref{fig:spec}C).
For each pair of values of $\nu_\mathrm{p}$ and $F_\mathrm{p}$, we did an equipartition analysis for relativistic targets as in \cite{Barniol2013,Matsumoto2023}.
With this, at each epoch we estimated the size of the emitter, $R_\mathrm{eq}$, and obtained values ranging from $\approx 1.1\times 10^{16}$~cm to $\approx 3.6 \times 10^{16}$~cm. Converting this to angular scales, we find it is indeed $\ll 0.2~\rm mas$, consistent with the detection of only a point-like structure in the VLBA data.

\section{Host galaxy properties}\label{sec:host_galaxy}
\noindent To constrain the host galaxy properties and estimate its contribution to the UVOT and VT detection, we used the Python package \texttt{CIGALE} for SED fitting. We gathered archival observations covering wavelengths from the UV to the mid-IR. For the mid-IR, we utilized W1 and W2 magnitudes from the AllWISE Source Catalog. For the near-IR, we obtained J, H, and K$_s$ magnitudes from the 2MASS All-Sky Point Source Catalog. For the optical bands, we calculated the mean g, r, i, z, and y magnitudes from the PS1 Detection table during the inactive period (MJD 55800--55880) of the host galaxy. For the UV, we extracted the NUV and FUV magnitudes from the All-sky Imaging Survey in the GALEX DR5 catalog.

We assumed a delayed star formation history (SFH) with an optional exponential burst component and adopted the single stellar population model of \cite{Bruzual2003} for stellar population synthesis. To model dust attenuation and emission, we employed the modified attenuation law from \cite{Calzetti2000} and dust emission models from \cite{Draine2014}. 
Overall, the fit yields a reduced $\chi^2$ of 0.26. To evaluate any AGN contribution to the system, we incorporated the AGN model from \cite{Fritz2006} into our model. However, this did not improve the fit, as the $\chi^2$ value remained unchanged, indicating that the AGN contribution to the SED is negligible. We show the best-fitting model in fig.~\ref{fig:host_info}B.

Galaxy properties and their uncertainties were estimated using the Bayesian-like method implemented in \texttt{CIGALE}. Eventually, the obtained total stellar mass of the galaxy, the stellar gas mass, the star formation rate, the stellar population age ($t_{\rm age}$), and the e-folding times of the populations ($\tau$) are $10^{9.57\pm0.26} M_\odot$, $10^{9.12\pm0.35} M_\odot$, $1.61\pm0.77$, $3.43\pm2.20$ Gyr, and $2.85\pm1.62$ Gyr, respectively.   
Based on the derived values of $t_{\rm age}$ and $\tau$, the SFH suggests an initial intense burst of star formation that rapidly peaked and has subsequently declined over a prolonged period, resulting in the current mature state of the galaxy. We also estimated the magnitudes of the host galaxy in the UVOT bands (V, B, U, UVW1, UVM2, and UVW2) to be 16.95, 17.43, 18.49, 19.03, 19.27, and 19.39, respectively, and in the VT bands (R and B) to be 16.29 and 16.98, respectively.

\section{Black hole mass estimation}\label{sec:bhmass}
\noindent We independently measured the black hole (BH) mass using the single-epoch method and the M-sigma relation.
These measurements were based on the medium-resolution DESI spectrum taken prior to the TDE. 

Before deriving the properties of the emission lines, it was crucial to accurately subtract the host galaxy's continuum and absorption features. To achieve this, we used the Penalized PiXel Fitting software (pPXF), originally developed by \cite{Cappellari2004} and further refined by \cite{Cappellari2017}. The pPXF incorporates stellar templates from the MILES Library\cite{Sanchez2006}. This library contains 985 well-calibrated stellar spectra covering a wavelength range of 3525–7500 \AA, with a spectral resolution characterized by a FWHM of 2.51 \AA~and a velocity dispersion ($\sigma$) of approximately 64 \kms.

The velocity dispersion ($\sigma = 118 \pm 46$ \kms) of \abw was determined by fitting the continuum and absorption features in the DESI spectrum using the pPXF. 
The flux at 5100 \AA~of the continuum was hence calculated as $3.6 \pm 1.0 \times 10^{-17}\rm~erg~s^{-1} A^{-1}$, following spectral decomposition method outlined by \cite{Du2024}. This leads to a luminosity of $L_{5100}\sim3.5 \pm 0.9 \times 10^{41}~\rm erg~s^{-1}$.
The FWHM of the H$\beta$ and H$\alpha$ broad emission lines were measured to be $3753\pm155$ \kms and $3298\pm81$ \kms by fitting the same spectrum with Gaussian functions after subtracting the continuum component.
The spectral decomposition reveals that the H$\beta$ emission line is weak and substantially affected by absorption from the host galaxy. The quoted uncertainty reflects only the formal fitting error and may underestimate the true uncertainty due to the low SNR.
Therefore, we applied the empirical relation between the FWHM of the H$\beta$ and H$\alpha$ emission lines with a rms scatter of 0.1 dex, as described by \cite{Greene2005}, to correct the FWHM of the H$\beta$ emission line. As a result, we obtained a corrected FWHM of $\sim3658\pm287$ \kms for H$\beta$. The error is estimated using Monte Carlo simulations, accounting for both measurement uncertainties and the intrinsic scatter of the relation.


Overall, adopting the corrected FWHM of the H$\beta$ line and $L_{5100}$, we derived a BH mass of $\log(M_{\bullet}/M_{\odot}) = 6.7 \pm 0.2$ using the single-epoch virial method. 
The random error is estimated using Monte Carlo simulations, taking into account measurement uncertainties, the virial factor $f$, and the scatter in the $R-L_{5100}$ relation.
The systematic error is determined by comparing the discrepancies among estimates derived from different empirical calibration relations.
Using the M-sigma relation established by \cite{Kormendy2013}, we obtained $\log(M_{\bullet}/M_{\odot}) = 7.4 \pm 0.7$, while the relation from \cite{McConnell2013} yielded $\log(M_{\bullet}/M_{\odot}) = 7.0 \pm 0.9$.
The final BH mass estimate based on $M-\sigma$ relation is determined to be $7.2 \pm 0.8\,(\rm stat.) \pm 0.4\,(\rm sys.)$.
Considering a systematic uncertainty of 0.5 dex associated with the single-epoch virial method (e.g., \cite{McLure2002, Vestergaard2006, Shen2015}), the BH masses derived from the two methods are consistent within uncertainties.
For further analysis in this work, we adopted $\log(M_{\bullet}/M_{\odot}) = 6.7\pm0.5$.

\section{Model setup and constraints}\label{sec:model}
The synchronized quasi-periodic flux variations observed in the X-ray and radio bands of \abw are naturally consistent with the theoretically predicted disk-jet Lense-Thirring (LT, \cite{LenseThirring1918}) precession model. The blackbody-dominated X-ray spectrum and the co-variation between the disk temperature and the X-ray luminosity further support this scenario.

In TDEs, the nascent accretion disk and the associated jet are highly likely to be misaligned with the spinning BH's equatorial plane, resulting in LT precession due to spacetime frame-dragging.
Following the disruption of a solar-type star by a SMBH with $M_{\bullet}\sim10^{6.7\pm0.5}~M_\odot$, the accretion rate is expected to exceed the Eddington limit during the first year. 
During this phase, the disk is geometrically thick with the semi-thickness, $H/R\geq \alpha$, where $H$ and $R$ is the disk's height and radius, respectively, and $\alpha$ is the viscosity parameter \cite{Shakura1973}. Under these conditions, the disk undergoes rigid-body precession, moving as a single unit.

In the super-Eddington accretion regime of TDEs, the thermal luminosity is expected to remain roughly constant as the accretion rate declines \cite{Krolik2012}. The two plateaus observed in the X-ray lightcurve of AT2020afhd prior to day 300 likely indicate periods of sustained super-Eddington or Eddington-limited accretion.
Overall, we adopt a commonly used TDE disk model described in \cite{Strubbe2009,Franchini2016}, where the disk remains geometrically thick with $H/R \sim 1$. Throughout, we assume that the jet is launched perpendicular to the plane of the disk.

\subsection{X-ray variations and black hole spin estimation}\label{sec:disk_precession}
To test whether the quasi-periodic variations in X-rays could be attributed to disk LT precession, we investigated the X-ray flux as a function of the angle $\psi(t) = \cos ^{ -1}(\hat {r}_{\rm{obs}} \cdot \hat {l}_{\rm{disk}}(t) )$, which represents the angle between the observer’s line of sight (the direction is denoted by the unit vector $\hat {r}_{\rm{obs}}$) and the disk’s angular momentum (the direction at time $t$ is denoted by the unit vector $\hat{l}_{\rm{disk}} (t)$). $\hat {r}_{\rm{obs}}$ is parameterized by the viewing angle $\theta_{\rm obs}$ (with respect to the BH spin axis $Z$) and $\phi_{\rm obs}$ as shown in Fig.~\ref{fig:schematics_model}A. The disk precesses with an angle $\theta_{\rm i}$ around the BH spin axis. Consequently, $\hat {l}_{\rm{disk}}(t)$, and thus $\psi (t)$, evolved over time due to LT precession.

Following \cite{Strubbe2009,Franchini2016}, the accretion disk can be modeled as a multicolor blackbody, with the effective temperature given by:
\begin{eqnarray}
 && T_{\rm eff,disk} = \left( \frac{3 G M_\bullet \dot{M} f_{\rm g}}{8\pi \sigma R^3} \right)^{1/4}  \left\lbrace  \frac{1}{2} + \left[\frac{1}{4} +\frac{3}{2} f_{\rm g} \left(\frac{\dot{m}}{ \epsilon} \right)^2 \left(\frac{r}{2}\right)^{-2} \right]^{1//2} \right\rbrace^{-1/4}, 
\end{eqnarray}
where $r = R/R_{\rm g}$, with $R_{\rm{g}}=G M_{\bullet}/c^2$ being the gravitational radius of the BH, $f_{\rm g} = 1-(R_{\rm ms}/R)^{1/2}$, with $R_{\rm ms}$ being the innermost stable circular orbit (ISCO), and $\epsilon=1-E_{\rm ms}$ is the radiative efficiency of disk, with $E_{\rm ms}$ being the specific energy at $R_{\rm ms}$ \cite{Bardeen1972}. $R_{\rm ms}$ and $E_{\rm ms}$ are functions of the BH spin parameter $a_\bullet$, which can be constrained using the LT precession period (see also Fig.~\ref{fig:MCMC_precession}A). The parameter $\dot{m}$ is the disk mass accretion rate $\dot{M}$ normalized to the Eddington accretion rate, $\dot{M}_{\rm Edd}=L_{\rm Edd}/c^2/\epsilon$. Here, $L_{\rm Edd}=1.26\times 10^{44} M_{\bullet,6} \ \rm erg \ s^{-1}$ is the Eddington luminosity, with $M_{\bullet,6}$ representing the BH mass scaled by $10^6 M_\odot$. 
In the super-Eddington phase, the disk thickness increases with the radius, reaching a maximum $H/R\sim 1$ near the disk outer radius $R_{\rm out}$ \cite{Franchini2016}. 
In our model, we set $R_{\rm out}$ to the circularization radius $R_{\rm cir}$ of the debris, such that $R_{\rm out}=R_{\rm cir}\simeq 2 R_{\rm tde}$, assuming the impact parameter $\beta=R_{\rm tde}/R_{\rm p} \simeq 1$ \cite{Franchini2016}. Here, $R_{\rm{tde}} \simeq (M_{\bullet}/M_*)^{1/3} R_*$ represents the tidal disruption radius, where $M_*$ and $R_*$ are the mass and radius of the disrupted star, and $R_{\rm p}$ is the pericentre of its orbit. 
Therefore, at large $\psi$, the X-ray emission from the inner disk is obscured by the outer, thicker portion of the disk, resulting in the observed sharp dips in the X-ray lightcurve. The X-ray variations are attributed to changes in the disk's projected area and periodic obscuration by the outer disk. For the same reason, the disk temperature should also modulate over the rigid-body precession period (observe the high-temperature inner part of the disk/X-ray peak when viewed at small $\psi$, and the low-temperature outer part/X-ray dip at large $\psi$), which is consistent with the observations in \abw (Fig.~\ref{fig:spec}A). This supports the precession of an extended disk rather than a narrow ring \cite{Pasham2024}.

Using the MCMC approach \cite{emcee}, we fitted the 0.1--2 keV X-ray lightcurve, observed between August 3 and October 21, 2024, where the X-ray quasi-periodic variations were clearly detected, with the proposed model. 
The model assumes a disk thickness ratio of $H/R \sim 1$ and an accretion rate of $\dot{m}\sim 1$. 
Flat priors were applied, including $0^\circ < \theta_{\rm i} < 30^\circ$ for the disk precession angle, $0^\circ < \theta_{\rm obs} < 90^\circ$ for the observer’s inclination/viewing angle, $0^\circ < \phi_{\rm obs} < 360^\circ$ for the azimuthal angle between the disk and the line of sight, and the variation amplitude constrained to exceed one order of magnitude.
We initialized 30 walkers to explore the parameter space, each taking 10,000 steps to solve for the maximum likelihood solution. The best-fitting parameters were $\theta_{\rm obs} \sim 38.4 ^{+0.5\circ}_{-0.6}$, $\theta_{\rm i} \sim 14.5^\circ \pm0.5^\circ $, and $\phi_{\rm obs} \sim 188.4^\circ \pm1.5^\circ$, as shown in Fig.~\ref{fig:MCMC_precession}B.

An important application of LT precession is to constrain the BH spin parameter.
Assuming a power-law surface density profile for the disk, $\Sigma(R)\propto R^{-p}$, the precession period can be expressed as \cite{Fragile2007,Stone2012,Franchini2016}
\begin{equation}
    T_{\rm prec} =\frac{2\pi}{\Omega_{\rm p}}= \frac{\pi G M_\bullet r_{\rm ms}^3}{a_\bullet c^3}  /(\frac{1-x_{\rm out}^{-1/2-p}}{x_{\rm out}^{5/2-p}-1} \frac{5/2-p}{p+1/2}),
\end{equation}
where $\Omega_{\rm p}$ is the precession frequency, $x_{\rm out}=R_{\rm out}/R_{\rm ms}$ and $r_{\rm ms} = R_{\rm ms} /R_{\rm g}$. 
Using power-law indices for a surface density profile of $p=0, 3/5, 3/4$ \cite{Franchini2016}, a SMBH mass of $M_\bullet=10^{6.7\pm0.5}~M_\odot$ and assuming the disruption of a solar-like star, we found that $a_\bullet$ fell within the range of $-(0.46-0.14)$ or $0.11-0.35$ (Fig.~\ref{fig:MCMC_precession}A).
A negative spin parameter would have resulted in a larger innermost stable circular orbit, potentially suppressing the high-energy X-ray emission from the accretion disk. Therefore, a positive spin parameter was favored.

Approximately 300 days after the re-brightening, the 19.6-day quasi-periodic variations disappeared, and the correlation between the X-ray and radio emissions broke down, suggesting that the angular momenta of the accretion disk and the SMBH may have aligned by that time.
Assuming a 300-day alignment timescale and adopting the system parameters derived above, one can place constraints on the disk viscosity parameter, $\alpha$. To this end, we consider two possible alignment mechanisms: viscous dissipation and disk cooling \cite{Bate2000, Franchini2016}. In both scenarios, $\alpha$ is allowed to vary freely within the range 0.01 to 1.
When viscous dissipation dominates the alignment \cite{Bate2000,Franchini2016}, the alignment timescale can be expressed as:
\begin{equation}
   t_{\rm align} \sim (1/\alpha) (H/R)^2 (\Omega/\Omega_{\rm p}^2),
\end{equation}
where $\Omega=c^3/(GM_\bullet)(1/(r^{3/2}+a_\bullet))$ represents the disk angular velocity. Using this relation yields $\alpha \sim 0.6$. Notably, the choice of $a_\bullet$ does not substantially affect this result.
Alternatively, if disk cooling drives the alignment \cite{Stone2012,Franchini2016}, the timescale can be estimated as:
\begin{equation}
t_{\rm thin} = t_{\min} \alpha^{-3/5}
\left[ 5\, f_{\rm g} \frac{M_*}{\dot{M}_{\rm Edd} t_{\min}} \frac{1}{r} \right]^{3/5},
\end{equation}
where $t_{\min} = 41 M_{\bullet,6}^{1/2} (M_*/M_{\odot})^{-1} (R_*/R_{\odot})^{3/2}$\,days represents the timescale for the disrupted material first coming back to pericentre. Adopting $a_{\bullet} \sim 0.11-0.35$ yields $\alpha \sim 0.2-0.4$.
In principle, the actual alignment timescale could exceed 300 days if the accretion disk had begun precessing prior to the observed re-brightening. In such a case, smaller values of $\alpha$ would be required under both alignment mechanisms.

\subsection{Radio variations}\label{sec:jet_precession}
The strong variations in radio flux and the high brightness temperature ($\sim$10$^{7}$ K) suggest that the radio emission originates from a compact outflow associated with BH activity, rather than star formation \cite{panessa19}. A roughly isotropic wind is unlikely to produce the strong variations observed in the radio emission. A natural inference is that the outflow is a relativistic jet with a bulk Lorentz factor $\Gamma$. A dip in the lightcurve occurred when jet moves away from the line of sight. 

For an observer at $\psi$ relative to the jet axis (as shown in Fig.~\ref{fig:schematics_model}A), the observed flux could be expressed as \cite{Granot2002,Lei2013}:
\begin{equation} 
F(\psi,t) = a_{\rm off}^4(\psi,\Gamma) F(0,a_{\rm off} t),
\end{equation}
where $a_{\rm off}(\psi,\Gamma)={\mathcal D}_{\rm off}/{\mathcal D}_{\rm on}=
(1-\beta_{\rm j})(1-\beta_{\rm j} \cos \psi)$ is the ratio of the on-beam to the off-beam Doppler factor, and $\beta_{\rm j}=\sqrt{1-1/\Gamma^2}$. For a TDE, the long-term lightcurve is defined by the fallback accretion rate, which is proportional to $t^{-5/3}$, and therefore have $F(\psi,t) = a_{\rm off}^{7/3}(\psi,\Gamma) F(0, t)$. The radio variations exhibited a peak-to-dip ratio, $F(\psi_{\rm peak})/F(\psi_{\rm dip})$, ranging from 1.9 to 4.3.
Using $\theta_{\rm obs} \sim 38.4^\circ$ and $\theta_{\rm i} \sim 14.5^{\circ}$, as suggested by the disk precession model, we determined $\psi_{\rm peak}= \theta_{\rm obs}-\theta_{\rm i} \simeq 23.9^\circ$ and $\psi_{\rm dip}= \theta_{\rm obs}+\theta_{\rm i}  \simeq 52.9^\circ$. From this, we constrained the jet's Lorentz factor to $1.2 \leq \Gamma \leq 1.6$.

Relativistic jets can be launched through either the Blandford-Znajek (BZ) mechanism \cite{Blandford1977} or the Blandford-Payne (BP) mechanism \cite{Blandford1982}. 
The BZ power is given by $L_{\rm BZ}=1.7 \times 10^{44}  a_{\bullet}^2 M_{\bullet,6}^2 B_{\bullet,6}^2 F(a_{\bullet}) \ {\rm erg \ s^{-1}} $, where the magnetic field strength threading the BH horizon is $B_{\bullet,6}=B_{\bullet}/10^{6} {\rm G}$, $F(a_\bullet)=[(1+q^2)/q^2][(q+1/q) \arctan q-1]$, and $q= a_{\bullet} /(1+\sqrt{1-a^2_{\bullet}})$ \cite{Lee2000,LeiZhang2011}. The peak radio luminosity could be approximated as $L_{\rm R}^{\rm peak} \sim L_{\rm BZ} a_{\rm off}^{7/3} (\psi_{\rm peak},\Gamma)$. Using $L_{\rm R}^{\rm peak} \sim 10^{38.4}\rm~erg \ s^{-1}$, a BH mass of $M_\bullet \sim 10^{6.7}~M_\odot$, BH spin $a_\bullet \sim 0.15$, $\psi_{\rm peak} \simeq 23.9^\circ$ and $\Gamma \sim 1.6$, the magnetic field was found to be $B_\bullet \sim 2.8 \times 10^3 $~G. For the BP mechanism, the jet power is given by $L_\mathrm{BP}=(B_\mathrm{ms}^\mathrm{p})^2R_\mathrm{ms}^4\Omega_\mathrm{ms}^2/32c$, where $B_\mathrm{ms}^{\rm{p}} = B_\bullet (R_\mathrm{ms}/R_\bullet)^{-5/4}$ and $\Omega_\mathrm{ms}$ is the angular velocity at $R_{\rm ms}$ \cite{Blandford1982,Livio1999}. Assuming the peak radio luminosity as $L_{\rm R}^{\rm peak} \sim L_{\rm BP} a_{\rm off}^{7/3} (\psi_{\rm peak},\Gamma)$, we derived a magnetic field of $B_\bullet \sim 1.5 \times 10^3$~G.

\clearpage
\renewcommand{\figurename}{Extended Data Figure}
\setcounter{figure}{0}
\renewcommand{\tablename}{Extended Data Table}
\setcounter{table}{0}

\clearpage
\begin{figure}[htp!]
\centering
\includegraphics[width=0.8\textwidth]{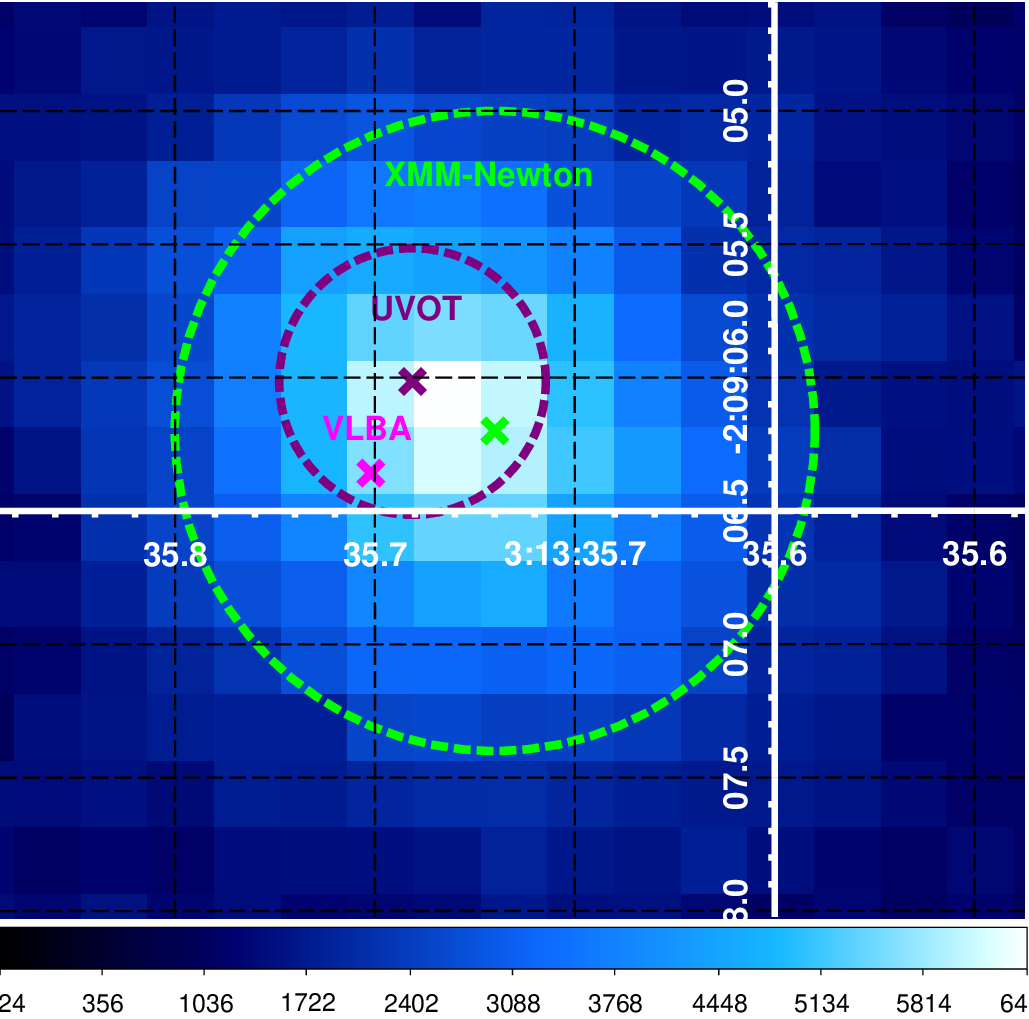}
\caption{\small \textbf{PanSTARRS-$g$ band image centered at the position of the galaxy LEDA~145386.} The colored crosses mark the centers of the multi-wavelength counterparts of \abw. The purple and green dashed circles, with radii of 0.5$^{\prime\prime}$ and 1.2$^{\prime\prime}$, respectively, indicate the positional uncertainties. The positional uncertainty for VLBA is too small to display.} \label{fig:image_multi}
\end{figure}

\begin{figure}[htbp!]
   \centering
	\includegraphics[width=0.49\textwidth]{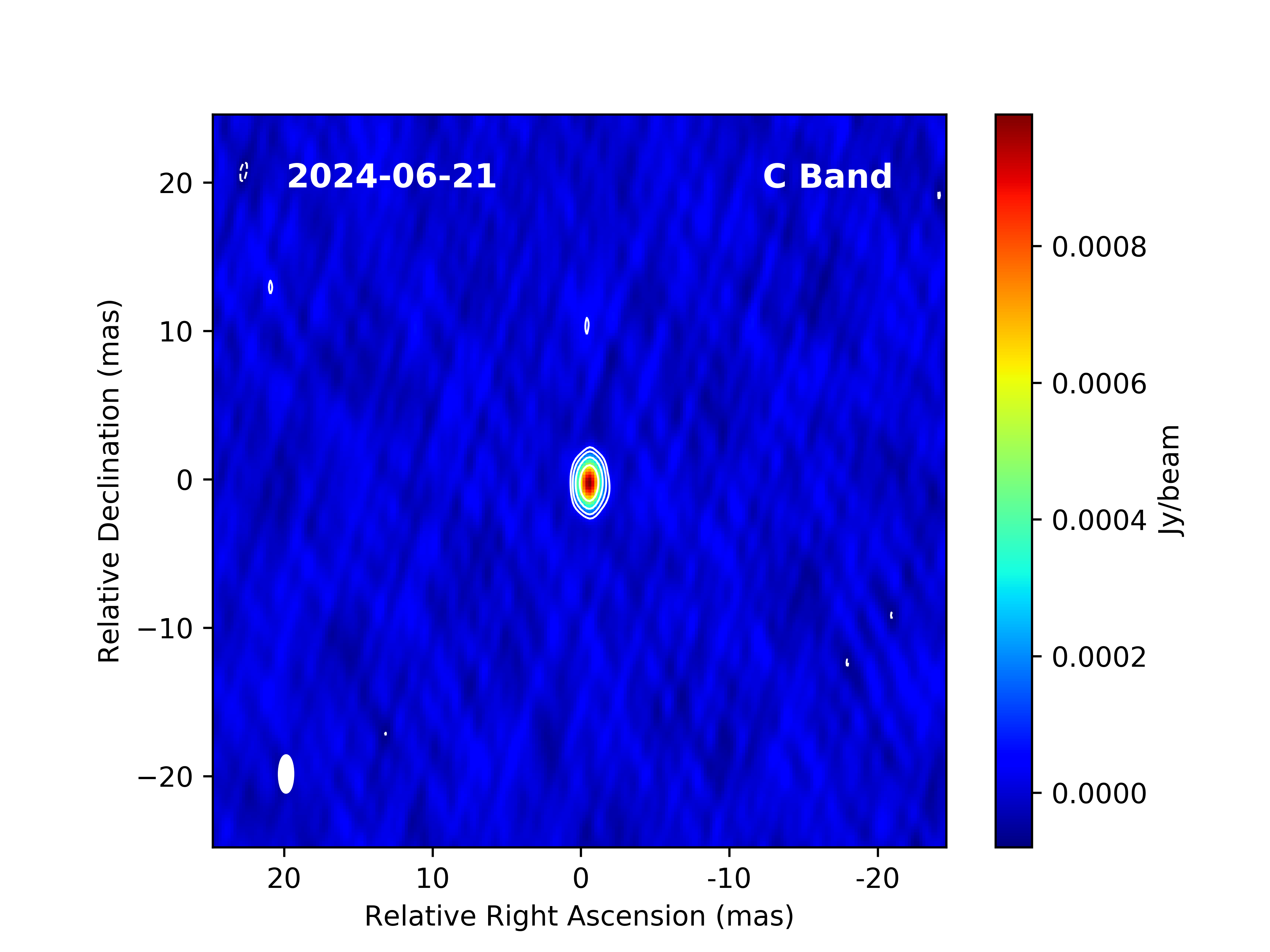}
    \hspace{-0.14cm}
 	\includegraphics[width=0.49\textwidth]{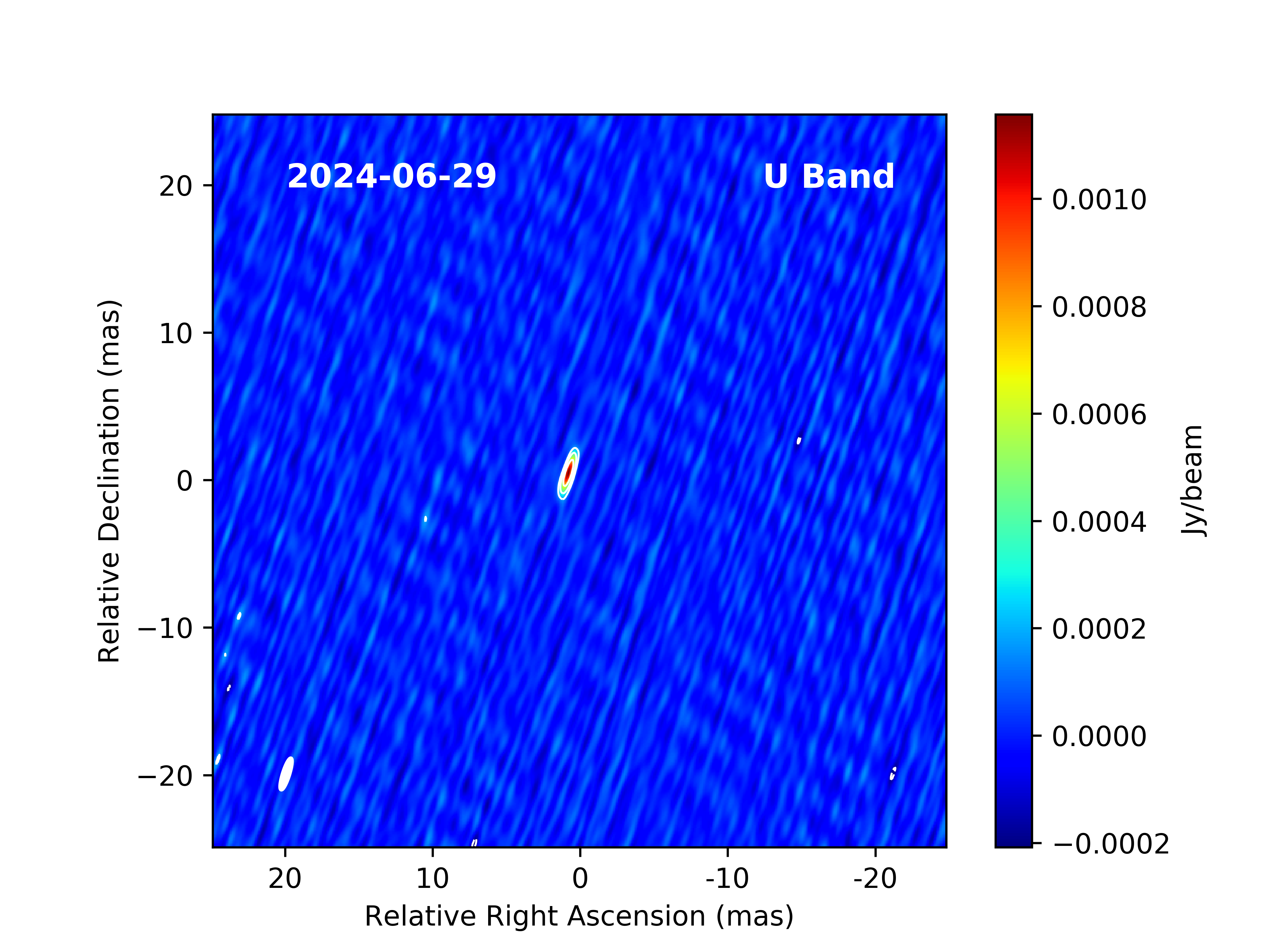}
	\includegraphics[width=0.49\textwidth]{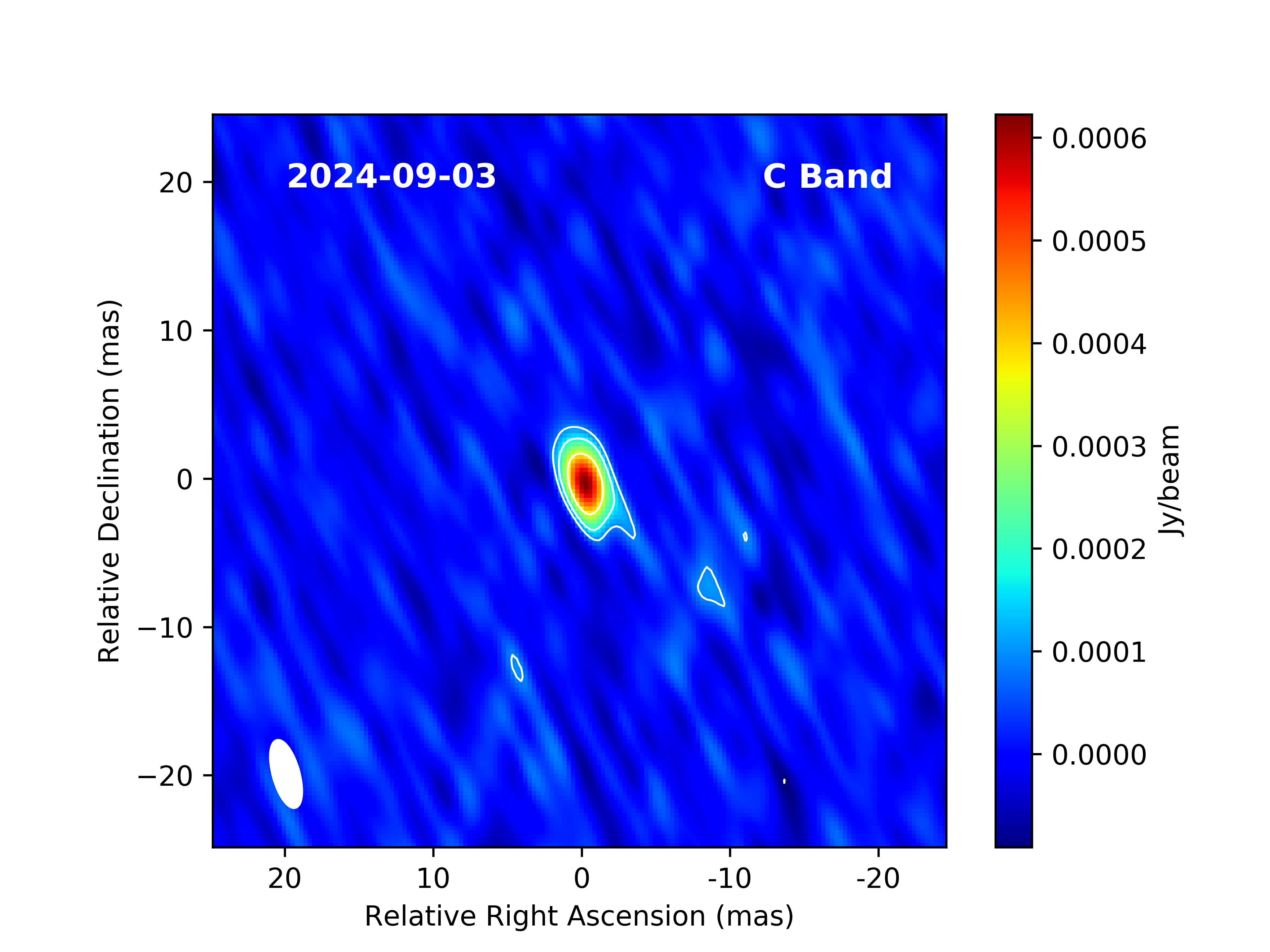}
    	\includegraphics[width=0.49\textwidth]{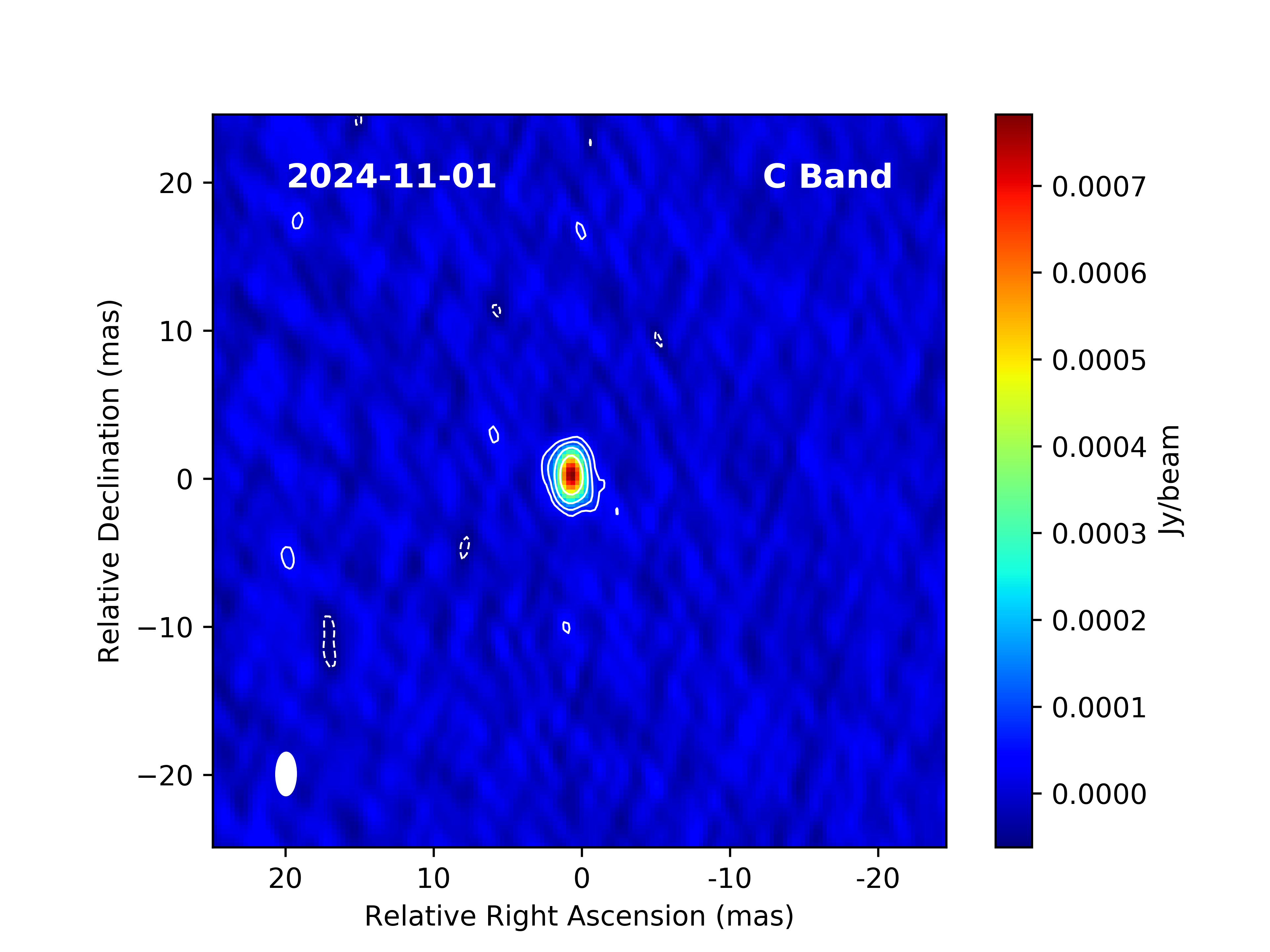}

        \caption{\textbf{VLBA maps in C and U band in four epochs.} In each map, the contours are drawn at ($-$1,1,2$^{n}$) times the off-source 3 $\times$ rms., which is 22, 58 and 28, 17 $\mu$Jy beam$^{-1}$ from top left to bottom right. The beam is displayed as white ellipse in the left bottom corner.} \label{fig:vlba}
\end{figure}

\begin{figure}[htbp!]
\centering
\includegraphics[width=1\textwidth]{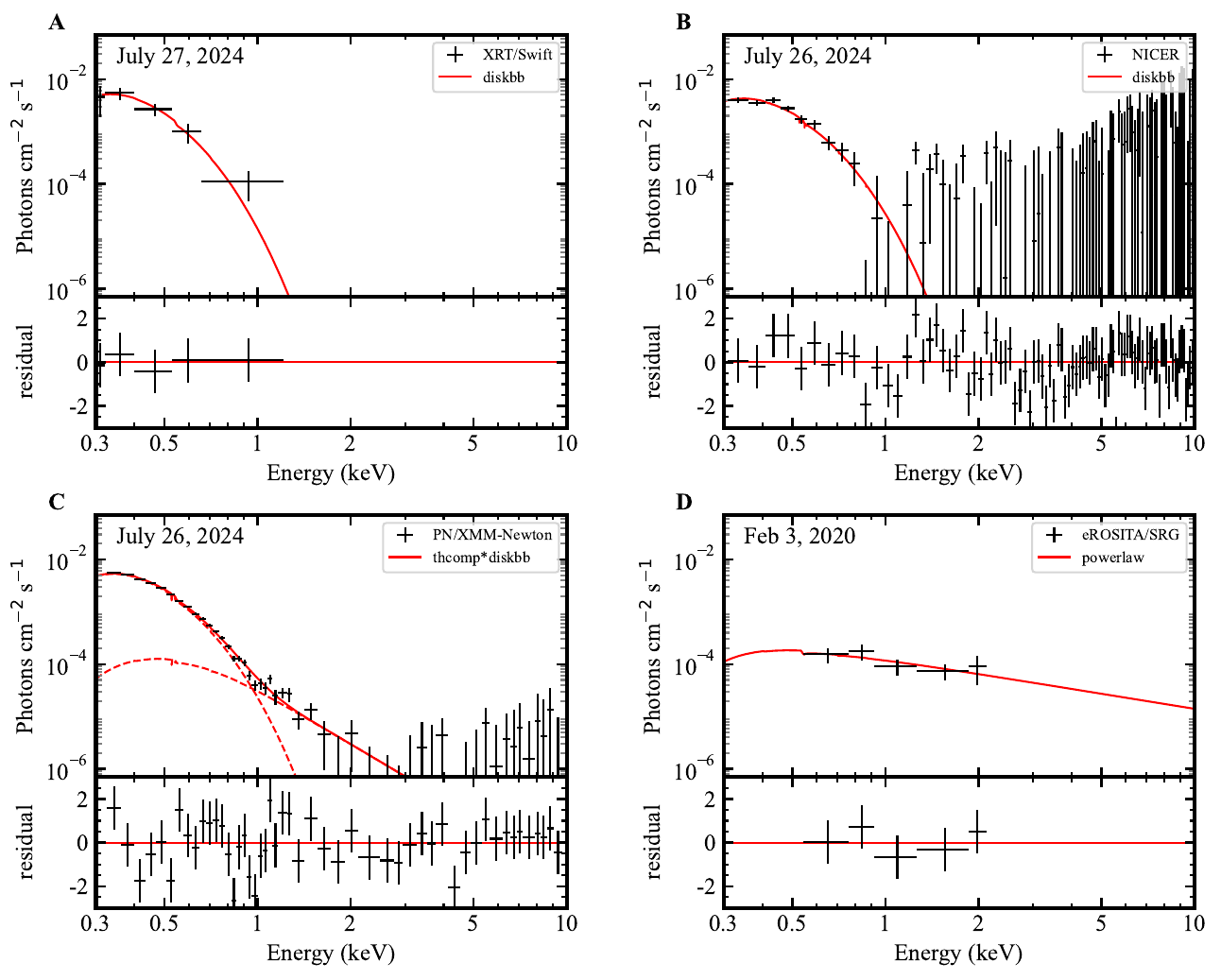}
\caption{\textbf{Illustration of four selected X-ray spectra.} (\textbf{A} to \textbf{C}) The XRT (ObsID: 00016492036), NICER (ObsID: 6705010126), and PN (ObsID: 0953010201) spectra were obtained within 1.2 days after the TDE, while (\textbf{D}) the eROSITA spectrum was taken prior to the TDE.}\label{fig:xray_spectra}
\end{figure}


\begin{figure}[htp!]
\centering
\includegraphics[width=0.7 \textwidth]{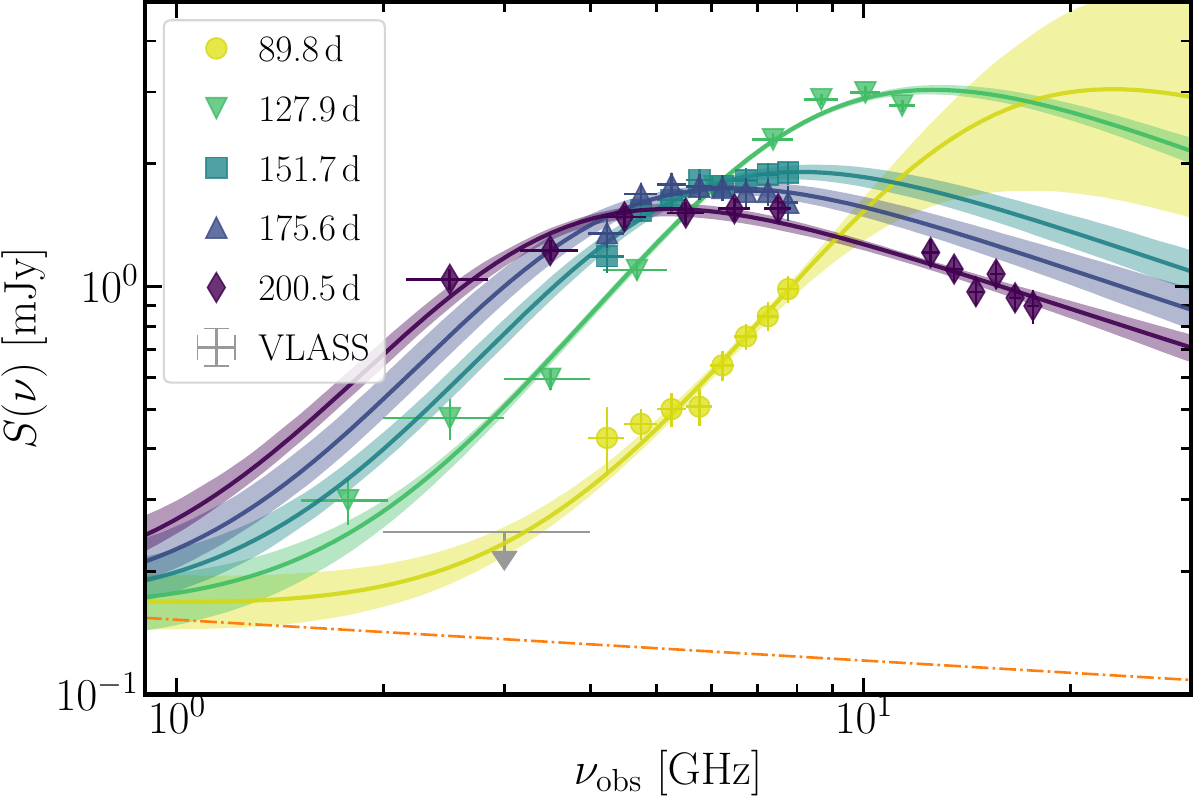} \hfill
\caption{\small \textbf{Multi-epoch radio SED using VLA data.} As the emitter expanded, it became optically thinner and the peak frequency decreased. The dot-dashed line represents the free-free emission from the host galaxy.} \label{fig:radio_sed}
\end{figure}

\begin{figure}[htbp!]
\centering
\includegraphics[width=1\textwidth]{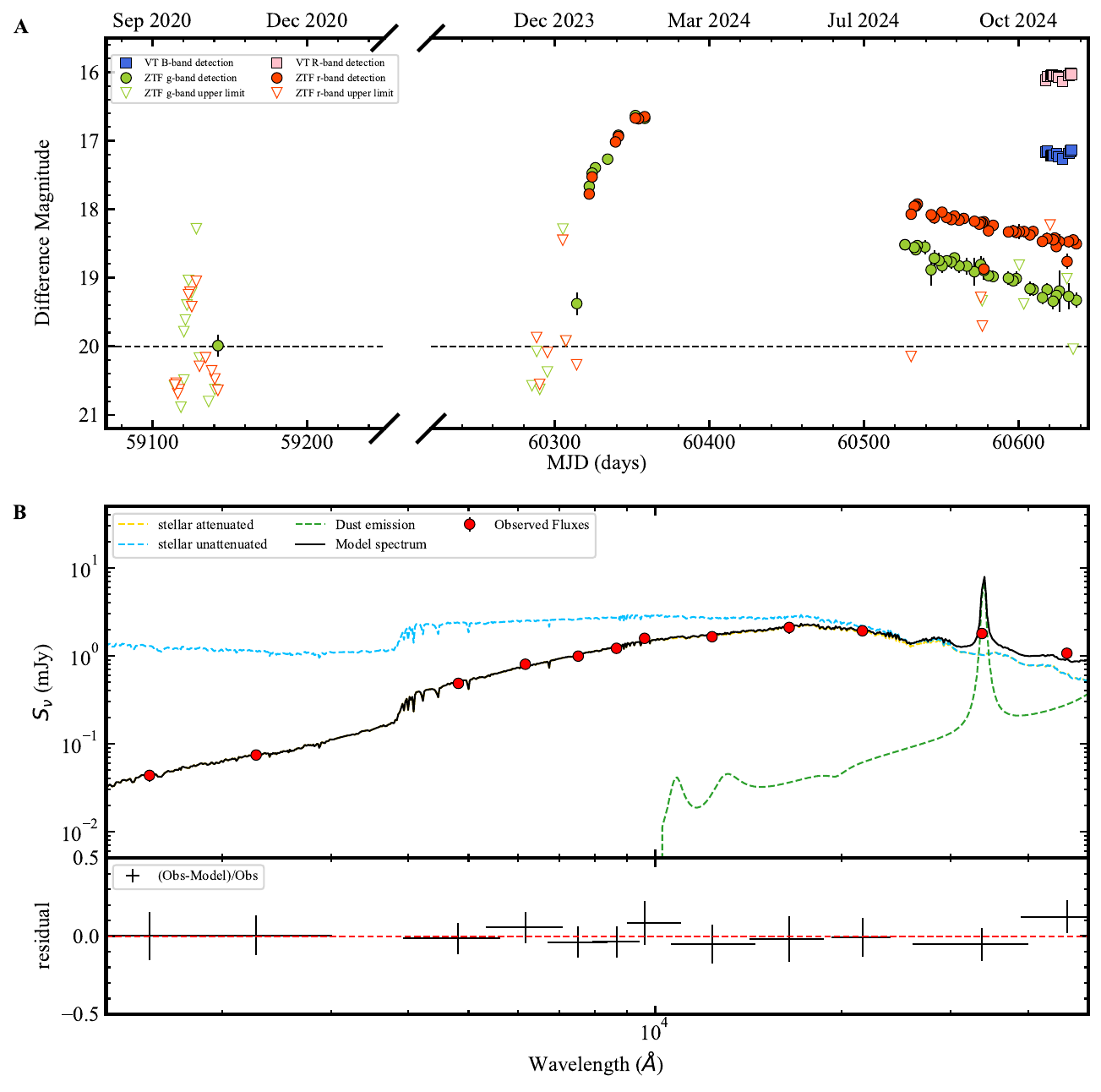}
\caption{\small \textbf{Some optical properties of AT2020afhd.} (\textbf{A}) ZTF and VT photometries. The ZTF data were corrected for extinction, while the VT data were corrected for Galactic
extinction and had the host contribution subtracted.
The dashed line indicates the magnitude level of the detection in 2020.
(\textbf{B}) Host-galaxy SED and the best-fitting model.}\label{fig:host_info}
\end{figure}



\clearpage


\clearpage

\end{bibunit}

\end{document}